\documentclass{aa}
\usepackage{graphicx,times}
\newlength{\fsize}
\renewcommand{\d}{\mathrm{d}}

\def\Real{{\rm I\mathchoice{\kern-0.70mm}{\kern-0.70mm}{\kern-0.65mm}%
  {\kern-0.50mm}R}}  
  \def\bx#1{\leavevmode\thinspace\hbox{\vrule\vtop{\vbox{\hrule\kern1pt
  \hbox{\vphantom{\tt/}\thinspace{\bf#1}\thinspace}}
  \kern1pt\hrule}\vrule}\thinspace}

\def\be{\begin{equation}} \def\ee{\end{equation}}

\begin{document}

\title{Cosmological Information from Quasar-Galaxy
  Correlations induced by Weak Lensing}
\author{Brice M\'enard\inst{1,2}\and Matthias Bartelmann\inst{1}}
\institute{$^1$ Max-Planck-Institut f\"ur Astrophysik, P.O.~Box 1317,
  D--85741 Garching, Germany\\
  $^2$ Institut d'Astrophysique de Paris, 98 bis Bld.~Arago, F--75014,
  Paris, France}
\date{\today}

\authorrunning{B.~M\'enard, M.~Bartelmann}
\titlerunning{QSO-Galaxy Correlations}

\date{Received / Accepted}

\abstract
  {The magnification bias of large-scale structures, combined with
   galaxy biasing, leads to a cross-correlation of distant quasars
   with foreground galaxies on angular scales of the order of arc minutes
   and larger. The amplitude and angular shape of the
   cross-correlation function $w_\mathrm{QG}$ contain information on
   cosmological parameters and the galaxy bias factor. While the
   existence of this cross-correlation has firmly been established,
   existing data did not allow an accurate measurement of
   $w_\mathrm{QG}$ yet, but wide area surveys like the Sloan Digital
   Sky Survey now provide an ideal database for measuring it. However,
   $w_\mathrm{QG}$ depends on several cosmological parameters and the
   galaxy bias factor. We study in detail the sensitivity of
   $w_\mathrm{QG}$ to these parameters and develop a strategy for
   using the data. We show that the parameter space can be reduced to
   the bias factor $\bar b$, $\Omega_0$ and $\sigma_8$, and compute
   the accuracy with which these parameters can be deduced from SDSS
   data. Under reasonable assumptions, it should be possible to reach
   relative accuracies of the order of $5\%$--$15\%$ for $\bar b$,
   $\Omega_0$, and $\sigma_8$. This method is complementary to other
   weak-lensing analyses based on cosmic shear.}

\maketitle

\section{Introduction}

Weak gravitational lensing by large-scale structures gives rise to a
magnification bias on distant quasar samples. Quasars behind matter
overdensities are magnified and therefore preferentially included into
flux-limited samples. Area magnification dilutes quasars on the sky
and counter-acts the magnification to some degree, but the number
count function of bright quasars is sufficiently steep to provide a
huge reservoir of faint quasars to be magnified above the flux
limit. Bright quasars in flux-limited samples thus occur
preferentially behind matter overdensities.

Galaxies are biased tracers of dark matter. If the bias is positive,
their number density is higher in matter overdensities. Since these
overdensities act as gravitational lenses for background sources, they
give rise to purely lensing-induced cross-correlations between
background quasars and foreground galaxies.

Typical angular scales for this effect range from a few arc minutes to
one degree. The lensing effect of individual galaxies does not matter
on such scales. Group- or cluster-sized haloes contribute to the
correlation only on scales below about one arc minute. On larger
scales, the expected signal is caused by the large-scale matter
distribution only. Measurements of the lensing-induced QSO-galaxy
correlation function therefore have the potential to constrain the
dark-matter power spectrum, several cosmological parameters, and the
galaxy bias.

Many such measurements have been undertaken (e.g.~Ben{\'\i}tez \&
Mart{\'\i}nez-Gonz\'alez 1995, Norman \& Impey 1999, Ben{\'\i}tez et
al.~2001, Norman \& Impey 2001). The {\em existence\/} of highly
significant large-scale QSO-galaxy correlations has been convincingly
demonstrated (cf.~Fugmann 1990; Bartelmann \& Schneider 1993, 1994;
Bartelmann et al.~1994; Rodrigues-Williams \& Hogan 1994; Ben{\^\i}tez
\& Mart{\^\i}nez-Gonz\'alez 1995; Norman \& Impey 1999), but the
measurement of the {\em amplitude\/} and {\em shape\/} of the
cross-correlation function is still highly uncertain (see Bartelmann
\& Schneider 2001 for a review). This is mainly because homogeneous,
sufficiently deep galaxy surveys of large portions of the sky have so
far been unavailable.

Upcoming wide-field surveys, above all the Sloan Digital Sky Survey
(SDSS; York et al. 2000), will provide huge, homogeneous samples of
quasars and galaxies covering a substantial fraction of the sky. The
detection and analyses of QSO-galaxy cross-correlations with high
signal-to-noise ratio is thus coming within reach. We investigate in
this paper which parameters can best be constrained with such a
measurement, and what accuracy we can expect to achieve.

The main route to extract cosmological information from weak lensing
has so far been the exploitation of cosmic shear. This approach
measures the gravitational tidal field of the dark matter between
background galaxies and the observer. Despite the many difficulties in
measuring exact shape parameters of faint galaxies, impressive results
have been obtained in the past year, demonstrating the power of the
method (Bacon et al.~2000, Hoekstra et al.~2001, Kaiser et al.~2000,
Van Waerbeke et al.~2001, Wittman et al.~2000). However, it depends
upon the crucial assumption that galaxy ellipticities are
intrinsically uncorrelated (see Heavens et al.~2000 and references
therein).

Quasar-galaxy cross-correlations induced by weak lensing provide a
complementary method for analysing weak lensing by large-scale
structures which does not depend on this assumption, and does not need
any shape measurement.  It therefore allows an independent and welcome
cross-check of the cosmic shear results and, in addition, it allows
the determination of the galaxy bias factor.

The ratio $\Omega_0/b$ can be constrained using the ratio
$w_\mathrm{QG}/w_\mathrm{GG}$, where $w_\mathrm{GG}$ is the angular
auto-correlation function of the foreground galaxies (Ben\'\i tez \&
Sanz 2000). However, this method neglects all information contained in
the amplitude and angular shape of $w_\mathrm{QG}$. Motivated by
recent cosmic shear results, we investigate here how well cosmological
parameters can be constrained if the full information provided by
quasar-galaxy cross-correlations is used.

In this paper, we briefly outline the lensing theory of QSO-galaxy
cross-correlations and investigate on which parameters the
cross-correlation function depends. We show that the dimensionality of
the parameter space can be substantially reduced. We then use
simulated measurements to investigate how accurately the remaining
parameters can be constrained.

\section{Theory}

The theory of the QSO-galaxy cross-correlation function caused by weak
lensing was introduced and developed in several earlier studies. We
can therefore be brief here and limit the discussion to the issues
required later on.

Let $n_\mathrm{Q}$ and $n_\mathrm{G}$ be the number densities on the
sky of quasars and galaxies, respectively. The QSO-galaxy cross
correlation function is then
\begin{equation}
  w_\mathrm{QG}(\theta)=\frac
    {\left\langle
     [n_\mathrm{Q}(\vec\phi)-\bar{n}_\mathrm{Q}]
     [n_\mathrm{G}(\vec\theta+\vec\phi)-\bar{n}_\mathrm{G}]
     \right\rangle}{\bar{n}_\mathrm{Q}\,\bar{n}_\mathrm{G}}\;,
\label{eq:1}
\end{equation}
where the average extends over all positions $\vec\phi$ and all
directions of $\vec\theta$. The average number densities of quasars
and galaxies are $\bar n_\mathrm{Q}$ and $\bar n_\mathrm{G}$,
respectively. If there is no overlap in redshift between the quasar
and galaxy populations, this correlation is exclusively due to
lensing, and can be written (Bartelmann, 1995)~:
\begin{equation}
  w_\mathrm{QG}(\theta)=2\,(\alpha-1)\,\bar b(\theta)\,
  w_{\kappa\delta}(\theta)\;,
\label{eq:2}
\end{equation}
provided that the magnification $\mu$ is weak, $|\mu-1|\ll1$. Here,
$\alpha$ is the logarithmic slope of the cumulative quasar
number-count function near the flux limit and
$w_{\kappa\delta}(\theta)$ is the cross-correlation function between
the lensing convergence $\kappa$ and a suitably weighted projected
density contrast $\bar\delta$. The convergence itself is a weighted
projection of the density contrast, so that the correlation function
$w_{\kappa\delta}(\theta)$ is straightforwardly related to the
projected dark-matter power spectrum.

Equation~\ref{eq:2} provides an operational definition of a mean
galaxy bias factor $\bar b(\theta)$ on a given angular scale
$\theta$. As defined, it averages over all galaxies considered in the
cross correlation at an angular separation $\theta$ from the nearest
quasar, and thus involves averaging over redshift and possibly also
morphological type, unless galaxies are selected according to their
morphology. Being linear in the galaxy number density, the QSO-galaxy
cross-correlation is insensitive to a possible stochasticity of the
bias. In practice, measurements will determine the $w_\mathrm{QG}$
within an angular range $R$ around $\theta$. We will abbreviate the
average bias factor within that interval as $\bar b_R$.

Of course, the relation of the bias factor $\bar b_R$ to commonly used
definitions of the galaxy bias factor depends on how the bias
depends on physical scale, redshift, and galaxy type. We will further
discuss this issue in Sect.~2.2 below.

\subsection{The $\kappa$-$\delta$ cross correlation}

We now describe briefly how the cross-correlation function
$w_{\kappa\delta}$ between lensing convergence and projected density
contrast can be evaluated. More detail can be found in Bartelmann
(1995), Dolag \& Bartelmann (1997), Sanz et al.~(1997) and Bartelmann
\& Schneider (2001).

Since both $\kappa$ and $\bar\delta$ are weighted line-of-sight
projections of the density contrast, we aim at an expression relating
$w_{\kappa\delta}$ to the matter power spectrum.

Specifically, we can write $\kappa$ as the projection
\begin{equation}
  \kappa(\vec\theta)= \int_0^{w_\mathrm{H}}\d w\,
  p_\kappa(w)\,\delta[f_K(w)\vec\theta,w]\;,
\label{eq:3}
\end{equation}
where $w$ is the radial comoving distance along the line-of-sight,
$f_K(w)$ is the comoving angular diameter distance and $p_\kappa(w)$
is the projector
\begin{eqnarray}
  p_\kappa(w)&=&\frac{3}{2}\,\Omega_0~\left(\frac{H_0}{c}\right)^2
  \nonumber\\&\times&
  \int_w^{w_\mathrm{H}}\frac{\d w'}{a(w)}\,W_\mathrm{Q}(w')\,
  \frac{f_K(w)\,f_K(w-w')}{f_K(w')}\;.
\label{eq:4}
\end{eqnarray}
The ratio between the angular diameter distances $f_K$ represents the
usual effective lensing distance, $W_\mathrm{Q}(w)$ is the normalised
distance distribution of the sources, in our case the quasars, and
$a(w)$ is the cosmological scale factor.

Similarly, the projected density contrast $\bar\delta$ can be written
as
\begin{equation}
  \bar\delta(\vec\theta)=\int_0^\infty\d w\,p_\delta(w)\,
  \delta[f_K(w)\vec\theta,w]\;,
\label{eq:5}
\end{equation}
where $p_\delta(w)$ is the normalised distance distribution of the
galaxies that are cross-correlated with the quasars.

The correlation function $w_{\kappa\delta}$ can now be related to
the statistics of $\delta$,
\begin{eqnarray}
  w_{\kappa\delta}(\theta) &=&
  \langle\kappa(\vec\phi)~\bar\delta(\vec\phi+\vec\theta)\rangle
  \nonumber\\
  &=&\int\d w\,p_\kappa(w)\int\d w'\,p_\delta(w')\,\nonumber\\
  &\times&\langle
    \delta[f_K(w)\vec\theta,w]~\delta[f_K(w')(\vec\phi+\vec\theta),w']
  \rangle\;.\nonumber
\label{eq:6}
\end{eqnarray}
Using Limber's equation for the statistics of projected homogeneous
Gaussian random fields, inserting the Fourier transform of the density
contrast, and introducing the power spectrum $P_\delta(k)$, we find
\begin{eqnarray}
  w_{\kappa\delta}(\theta)&=&\int\d w\,
  \frac{p_\kappa(w)\,p_\delta(w)}{f_K^2(w)}\,\nonumber\\
   &\times&\int\frac{s\d s}{2\pi}\,
  P_\delta\left(\frac{s}{f_K(w)},w\right)\,\mathrm{J}_0(s\theta)\;.
\label{eq:7}
\end{eqnarray}
The quasar-galaxy cross-correlation function,
$w_\mathrm{QG}(\theta)=2\,(\alpha-1)\,\bar
b(\theta)\,w_{\kappa\delta}(\theta)$, describes the statistical excess
of quasars around galaxies with respect to a Poisson distribution, or
the excess of galaxies around QSOs, at an angular distance $\theta$.

\subsection{The bias factor}

The relation of the average bias factor $\bar b_R$ to the usual galaxy
bias factor depends on the properties of biasing and of the galaxies
considered. If biasing is linear and possibly stochastic, the bias
factor as defined here averages over $rb_\mathrm{var}$, where
$b_\mathrm{var}$ is the ratio of variances in the galaxy and
dark-matter distributions, and $r$ is the linear correlation
coefficient defined by Dekel \& Lahav (1999). For nonlinear biasing,
$\bar b_R$ averages over the biasing parameter $\hat b$, which is the
slope of the linear regression of the galaxy fluctuations on the
density contrast and whose relation to other conventional measures of
galaxy biasing becomes more involved (Dekel \& Lahav 1999). In other
words, the physical meaning of $\bar b_R$ is defined by
Eq.~(\ref{eq:2}), but its relation to other measurements of galaxy
biasing depends on the biasing model and remains to be specified if
such a relation needs to be established.

Generally, the galaxy bias can depend on time, scale, and galaxy
parameters like luminosity or morphological type. Numerical
simulations using semi-analytic galaxy-formation models predict how
the bias of different morphological galaxy types evolves with redshift
(cf.~Kauffmann et al.~1999), and such prescriptions can be used for
establishing the relation between $\bar b_R$ and other biasing
parameters. The impact of a possible scale dependence of the bias on
quasar-galaxy cross-correlations can also be taken into account. This
was studied by Guimar\~aes et al.~(2001), who used ratios between
measured and theoretical galaxy power spectra for estimating the mean
bias parameter in Fourier space, $\bar b(k)$. Naturally, such methods
strongly depend on the bias model.

For our application, the redshift dependence of the mean bias $\bar b$
can probably be neglected because the galaxies expected to be
cross-correlated with background quasars will be in a relatively
narrow redshift range peaking near $z\approx0.3$. Furthermore, the
scale dependence can be constrained.
Ben{\'\i}tez \& Sanz (2000) showed that the ratio
$w_\mathrm{QG}/w_\mathrm{GG}$ effectively measures $\Omega_0\,\bar
b/b_{\rm var}^2$,
without the need to know the cosmological parameters. For linear and
deterministic biasing, $b_{\rm var}$ reduces to $\bar b$ and we have~:
\begin{equation}
  \frac{w_\mathrm{QG}(\theta)}{w_\mathrm{GG}(\theta)}=
  Q\,\frac{\Omega_0}{\bar b(\theta)}
\label{eq_bias}
\end{equation}
for quasars and galaxies located at single redshifts $z_\mathrm{Q}$
and $z_\mathrm{G}$. Van Waerbeke (1998) extended this calculation to
realistic redshift distributions, for which the constant $Q$ becomes a
known scale-dependent function $Q(\theta)$, but remains only weakly
sensitive to cosmological parameters. Therefore, the method proposed
by Ben{\'\i}tez \& Sanz allows constraints on the scale dependence of
the bias.

We investigate here the possible constraints on cosmological
parameters that can be derived from the angular shape and amplitude of
$w_\mathrm{QG}$. For doing so, we assume that the bias is 
independent of scale on the angular scales we are considering,
i.e.~between $1'$ and $1^\circ$. A scale-dependent bias factor could
be taken into account assuming that the variation with scale can be
expressed by a smooth function $f(\theta)$, so that
\begin{equation}
  \bar b(\theta)=\bar b_R\,f(\theta)
\end{equation}
Models or measured constraints on
$f(\theta)$ can then be included into the calculation of
the cross-correlation function $w_\mathrm{QG}(\theta)$.

\subsection{Expected galaxy overdensity}

\begin{figure*}[th]
\includegraphics[width=0.45\hsize]{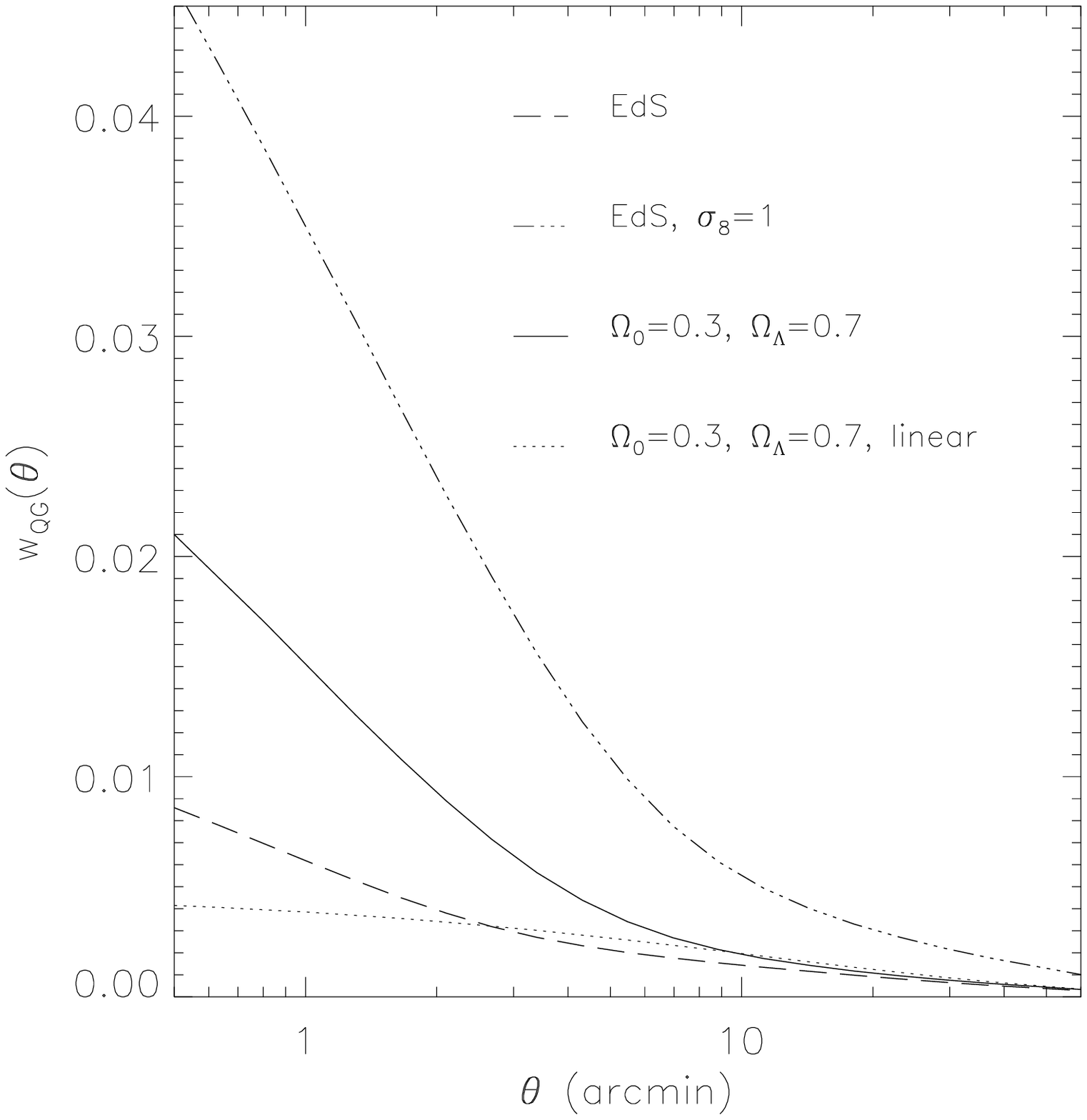}\hfill
\includegraphics[width=0.45\hsize]{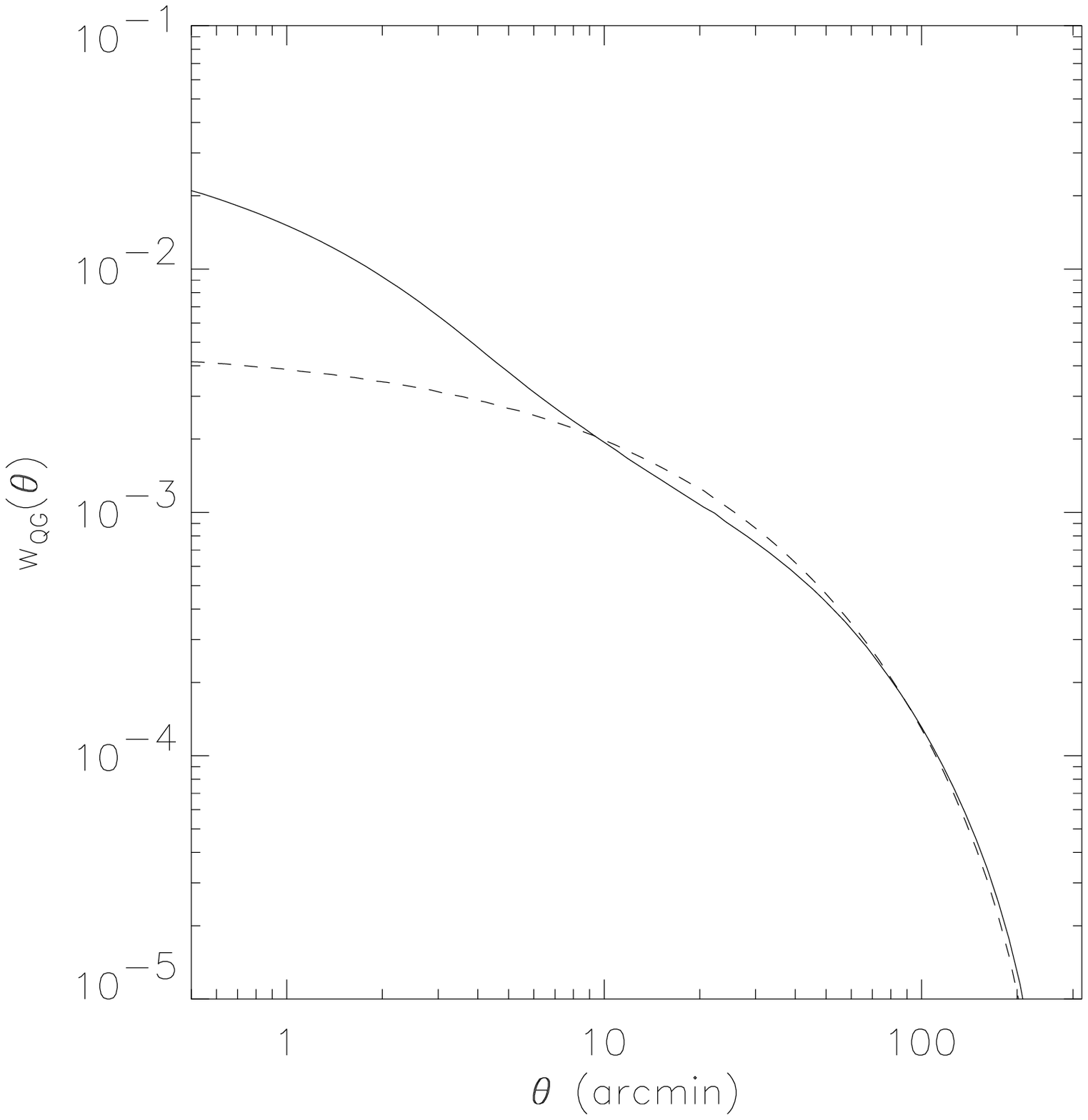}
\caption{{\em Left panel:\/} The QSO-galaxy cross-correlation function
  $w_\mathrm{QG}(\theta)$ is plotted for different cluster normalised
  cosmologies. A low-density flat universe with $\Omega_0=0.3$,
  $\Omega_\Lambda=0.7$ is plotted in solid line; an Einstein-de Sitter
  universe is plotted in dashed line, and in dashed-dotted line for a
  universe normalised with $\sigma_8=1$. Both curves were computed
  using the non-linearly evolving matter power spectrum. For
  comparison, the dotted line shows how the correlation function if we
  consider linear growth of the power spectrum for a $\Omega_0=0.3$,
  $\Omega_\Lambda=0.7$ cosmology. Here we have assumed $\alpha=2$, a
  bias factor of $\bar b=1$ and a shape parameter of $\Gamma=0.21$. The
  {\em right panel\/} shows how the correlation function drops at
  large scales before reaching negative values in a low-density flat
  universe.}
\label{shape}
\end{figure*}

We now evaluate Eq.~(\ref{eq:7}) under the following
assumptions. First, we choose a CDM power spectrum and use the
formalism by Peacock \& Dodds (1996) to approximate its non-linear
evolution. This is crucial because the linear approximation
underestimates the correlation amplitude on arc-minute scales by about
an order of magnitude (Dolag \& Bartelmann 1997). Unless specified
otherwise, we use the cluster normalisation constraint
$\sigma_8=0.52\,\Omega_0^{-0.52+0.13\,\Omega_0}$ from Eke et
al.~(1996).

Since we shall later need to compute the correlation function many
times to explore the parameter space, we assume that the redshift
distributions for source quasars and foreground galaxies are delta
functions, which speeds up the computation, but affects the amplitude
of $w_{\kappa\delta}(\theta)$ only by at most 20\% compared to
realistic distance distributions. We place all galaxies at redshift
$z=0.3$, and all quasars at redshift $z=1.5$.

Finally, we need a model for the quasar number counts as a function of
flux. In the quasar catalogue of the Sloan Early Data Release
(Schneider et al. 2001), the cumulative number-count function for
quasars brighter than $19^\mathrm{th}$ magnitude is well approximated
by a power law with a logarithmic slope of $\alpha\sim 2$.

Some illustrative results for $w_\mathrm{QG}(\theta)$ are presented in
Fig.~\ref{shape}. The curves in the left panel show the expected
cross-correlation functions for nonlinearly evolving density
perturbations in a flat, low-density universe ($\Omega_0=0.3$,
$\Omega_\Lambda=0.7$, cluster normalised with $\sigma_8=0.93$; solid
line), an Einstein-de-Sitter universe (cluster normalised with
$\sigma_8=0.52$, dashed line; and normalised to $\sigma_8=1$,
dashed-dotted line).

For comparison, the dotted curve was computed for an $\Omega_0=0.3$,
$\Omega_\Lambda=0.7$, cluster normalised cosmology, but assuming
linear growth of the density perturbations. At small angular scales
(below $\sim10'$), most of the correlation amplitude is contributed by
mildly nonlinear matter perturbations. The right panel shows the
correlation function out to large angular scales, where it drops below
zero. The break towards two degrees occurs when the first zero of the
Bessel function in Eq.~(\ref{eq:7}) moves across the physical scale of
the peak in the power spectrum. Hence, the correlation signal is
expected to drop sharply at these large angular scales.

Figure~\ref{shape} demonstrates that the quasar-galaxy
cross-correlation function varies substantially if the cosmological
model is changed. Therefore, it can be used as a tool to measure
cosmological parameters. We explore this issue in the next section.

\section{Dependence on cosmological parameters}

\subsection{Reduction of the parameter space}

Equation~(\ref{eq:7}) illustrates how the quasar-galaxy
cross-correlation function depends on various assumptions and
parameters. The redshift distributions of the source quasars and the
foreground galaxies enter through the functions $p_\kappa(w)$ and
$p_\delta(w)$, respectively. These distributions can be
observed. Likewise, the logarithmic slope $\alpha$ of the quasar
number-count function can directly be obtained through observations.

Therefore, the remaining unknowns of the cross-correlation function
are the dark-matter power spectrum and its growth, the cosmological
parameters, and the bias factor of the galaxies relative to the dark
matter. We now explore how sensitively the cross-correlation function
depends on these inputs.

First, we shall assume that the matter content in the Universe is
dominated by cold dark matter.  For a Universe containing pure cold
dark matter, the location of the peak of the power spectrum is defined
by $\Gamma=\Omega_0\,h$ (Bardeen et al.~1986), and it changes to
$\Gamma=\Omega_0\,h\,
\exp[-\Omega_\mathrm{b}\,(1+\sqrt{2\,h}\,/\,\Omega_0)]$ in presence of
baryons (Sugiyama 1995). For reasonably low values of the baryon
density parameter, the main effect of the preceding modification is
merely a weak change of the first relation between $\Gamma$ and
$\Omega_0$. For $\Omega_\mathrm{b}\,h^2=0.02$ (Wang et al. 2001), we
find that $\Gamma\simeq\,\Omega_0\,h-0.03$.  This detailed, weak
dependence on the baryon density does not qualitatively affect the
following results. We shall see in Sect.~4.2 how the parameter
estimation changes if we ignore any relation between $\Gamma$ and the
other parameters.

The remaining free parameters appearing in the quasar-galaxy
cross-correlation function are the cosmological parameters $\Omega_0$,
$\Omega_\Lambda$ and $h$, the normalisation of the power spectrum,
which is expressed by its variance $\sigma_8$ on scales of
$8\,\mbox{Mpc}/h$, and the bias factor $\bar b_R$.

\subsection{Independence on $\Omega_\Lambda$}

\begin{figure*}[ht]
  \includegraphics[width=0.3\hsize]{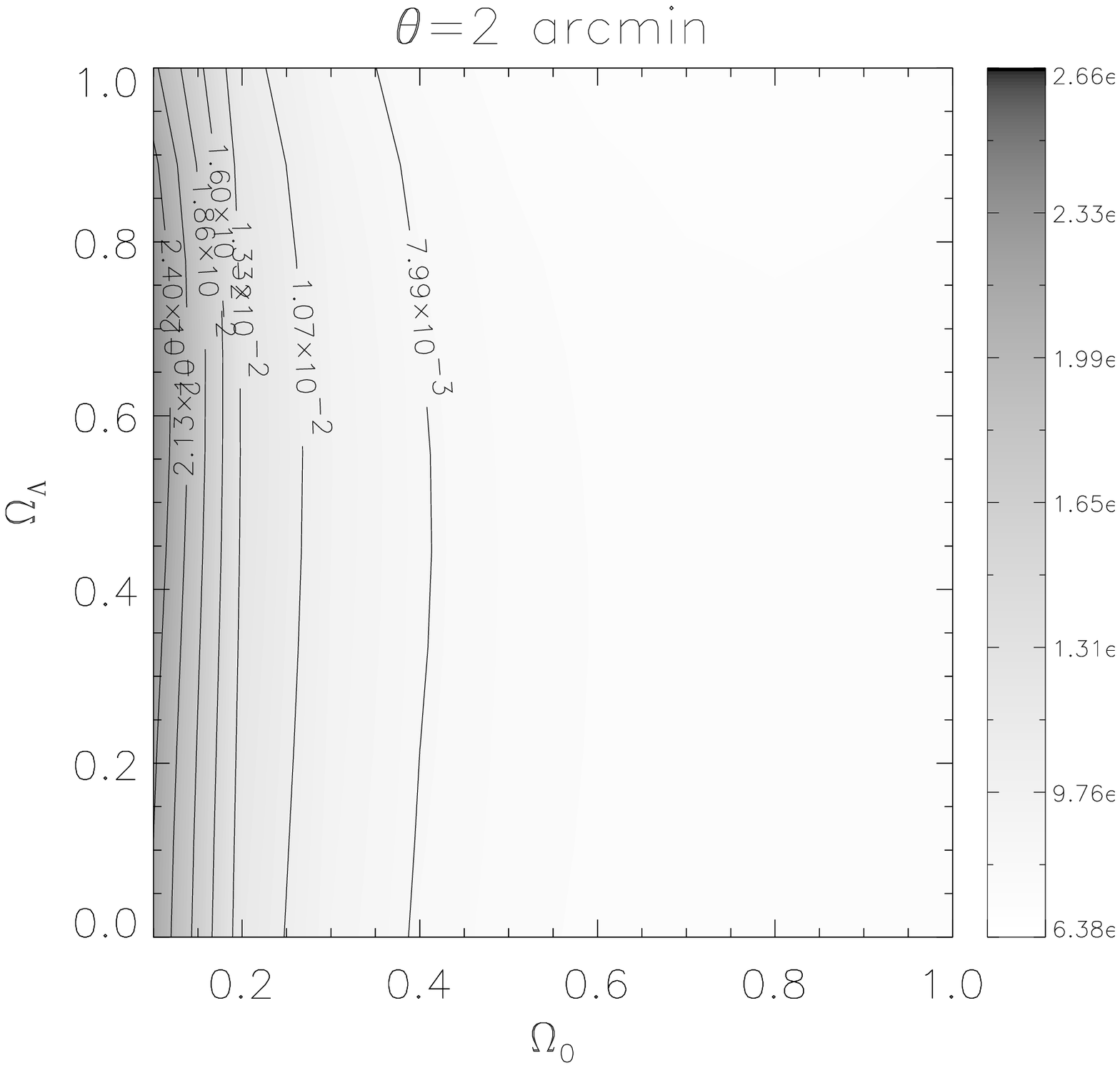}\hfill
  \includegraphics[width=0.3\hsize]{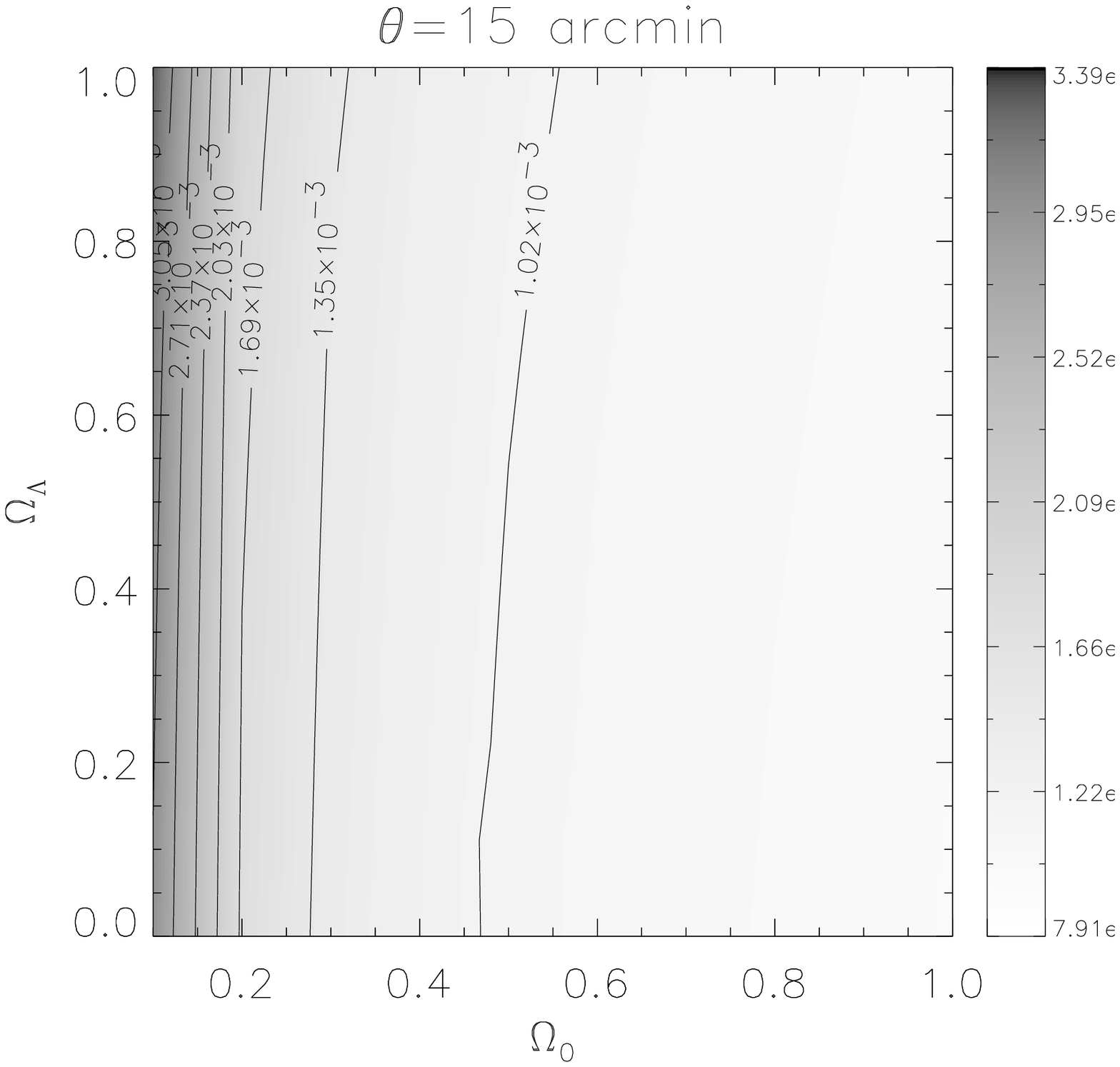}\hfill
  \includegraphics[width=0.3\hsize]{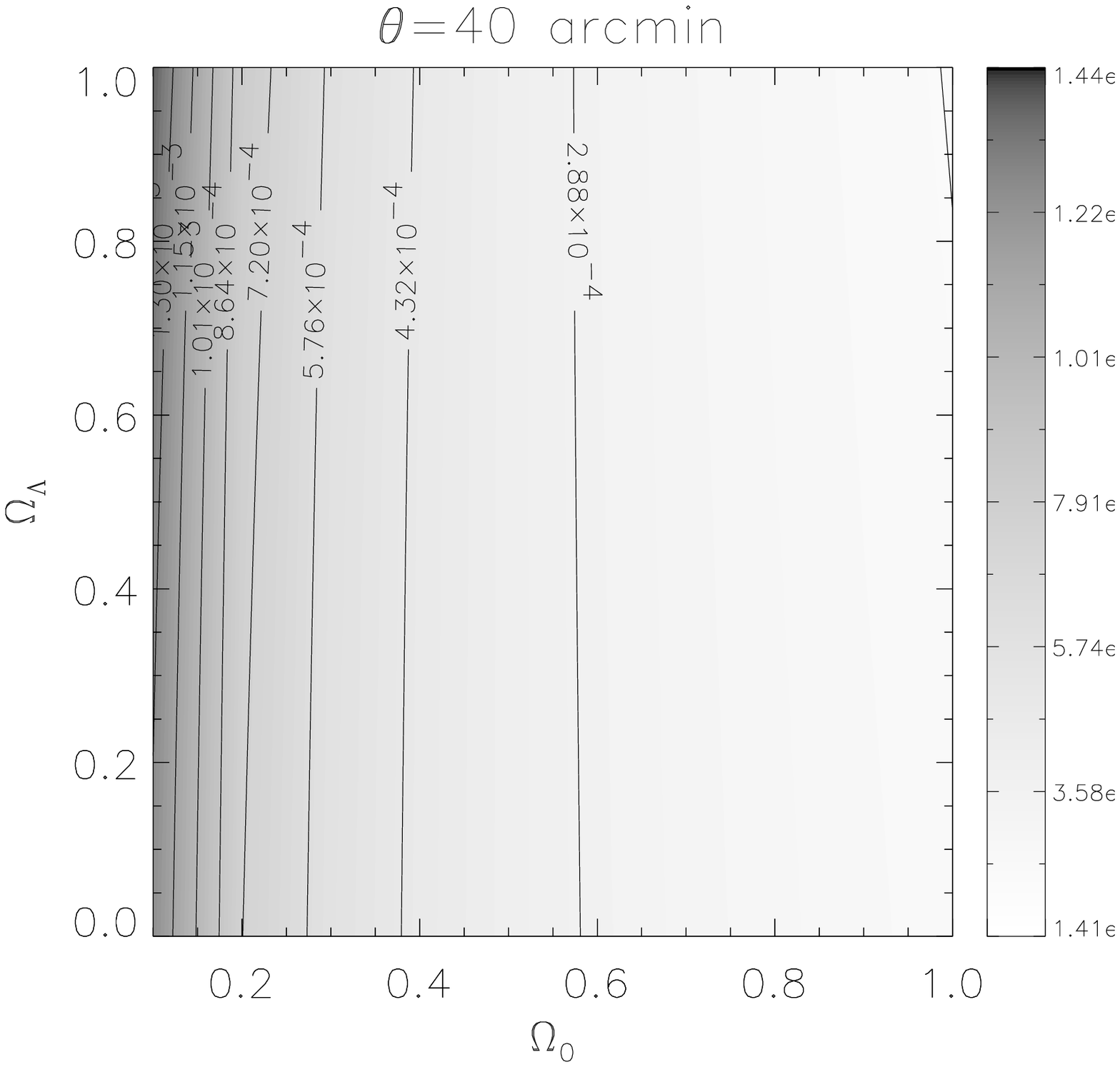}
\caption{The dependence of $w_{\kappa\delta}$ on $\Omega_0$ and
  $\Omega_\Lambda$ is shown for two different angular scales
  ($\theta=2'$, $15'$ and $40'$). Here we have used a cluster
  normalised cosmology, with $\Gamma=\Omega_0\,h$ and $\bar b=1$.  On
  scales greater than one arc minute, the $\Omega_\Lambda$-dependence
  is so weak that this parameter can safely be eliminated.}
\label{lambda_dependence}
\end{figure*}

Interestingly, $w_{\kappa\delta}(\theta)$ depends only very weakly on
the cosmological constant $\Omega_\Lambda$ on scales larger than one
arc minute. This is demonstrated by Fig.~\ref{lambda_dependence} for
different angular scales and different cosmologies. Above one arc
minute, the contours tend to align parallel to the $\Omega_\Lambda$
axis. At an angular scale of $15$ arc minutes for instance, the
variation of $w_{\kappa\delta}$ with $\Omega_\Lambda$ for a given
density parameter $\Omega_0$ is less than $10\%$.  For comparison,
increasing $\Omega_0$ from $0.1$ to $1$ changes the correlation
amplitude by roughly a factor of four.  Thus, focusing on
intermediate and large angular scales, the correlation function can be
considered insensitive to $\Omega_\Lambda$. This safely allows the
reduction of the effective parameter space to $\Omega_0$, $\sigma_8$,
$h$ and $\bar b_\mathrm{R}$.

\subsection{Dependence on $\Omega_0$ and $\sigma_8$}

Regarding the dependence of $w_{\kappa\delta}$ on $\Omega_0$ and
$\sigma_8$, a rough inspection of Eq.~(\ref{eq:7}) would give
$w_{\kappa\delta}\propto\Omega_0\,\sigma_8^2$, where the factor
$\Omega_0$ comes from Poisson's equation, and the scaling
$\propto\sigma_8^2$ reflects the normalisation of the power
spectrum. However, closer examination reveals a more complicated
dependence. Due to the nonlinear evolution of the power spectrum,
$\sigma_8$ does not only change the correlation amplitude, but also
its shape. Similarly, $\Omega_0$ affects the nonlinear evolution of
the power spectrum, and therefore also changes the shape of
$w_{\kappa\delta}$.

\begin{figure*}[ht]
  \includegraphics[width=0.3\hsize]{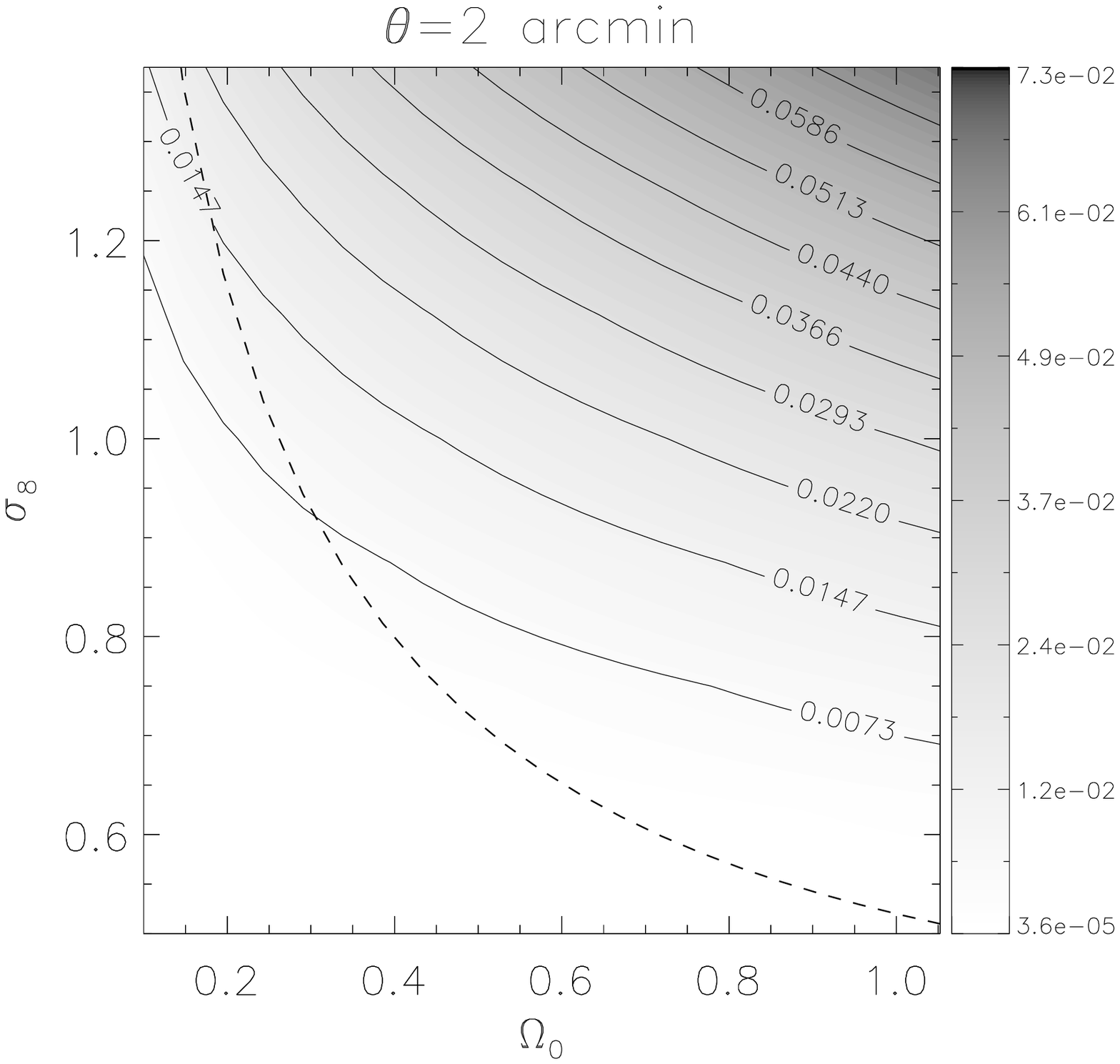}\hfill
  \includegraphics[width=0.3\hsize]{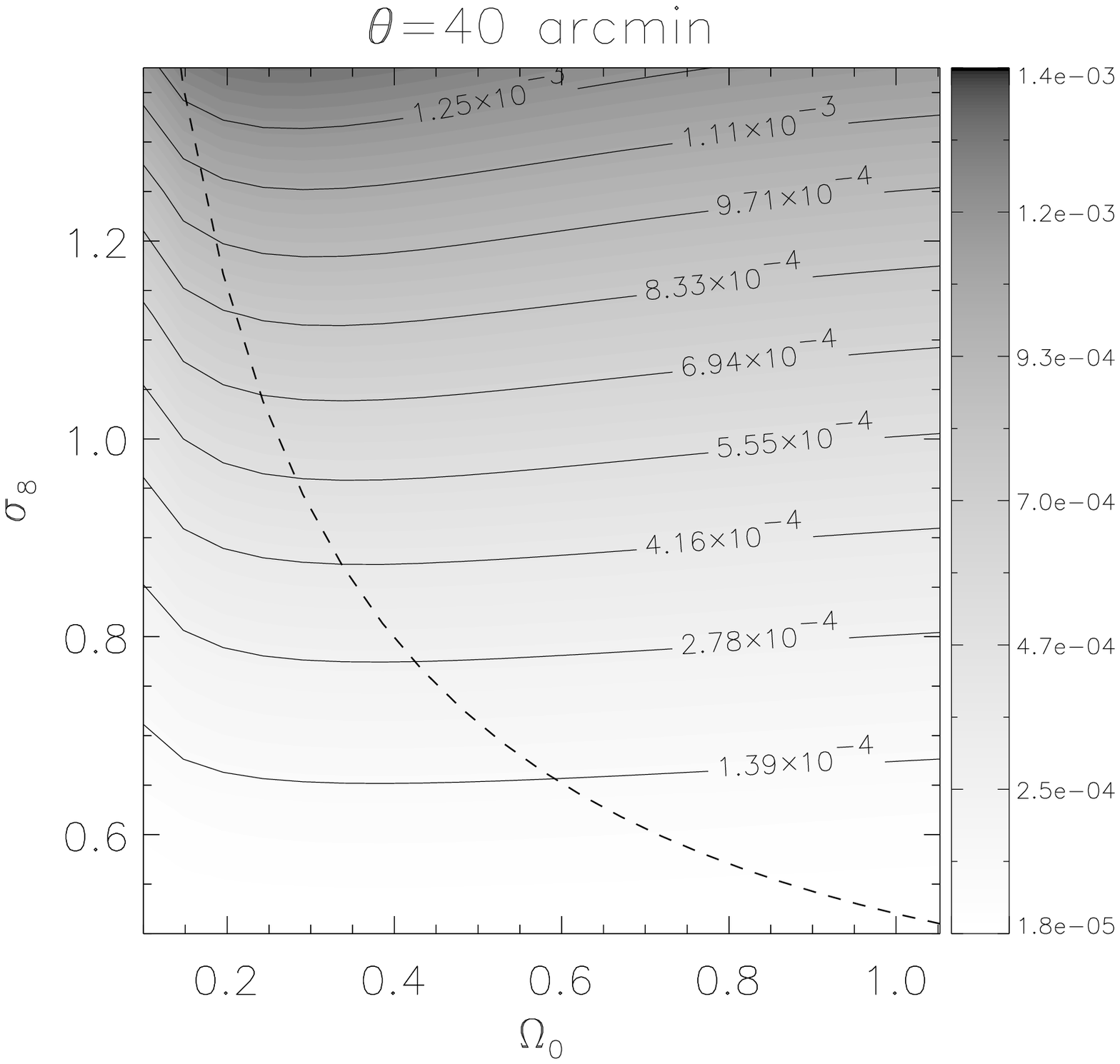}\hfill
  \includegraphics[width=0.3\hsize]{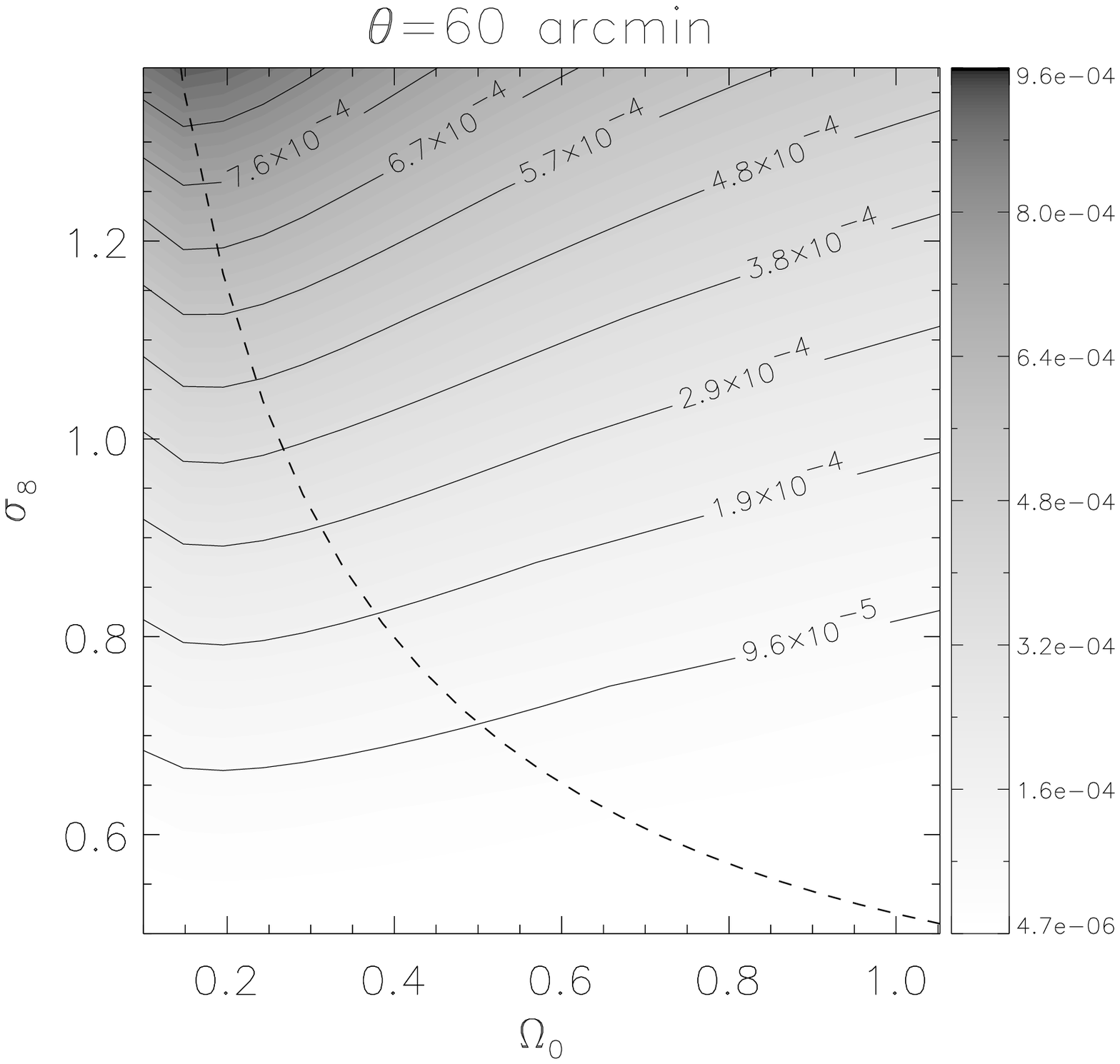}
\caption{The dependence of $w_{\kappa\delta}$ on $\Omega_0$ and
  $\sigma_8$ is shown for three different angular scales ($\theta=2'$,
  $40'$ and $60'$ from left to right) and assuming
  $\Gamma=\Omega_0\,h$.  We kept $h=0.72$ and $\bar b=1$ fixed for the
  plot. As the dependence on $\sigma_8$ and $\Omega_0$ changes with
  angular scale, the degeneracy between these two parameters can be
  broken. At $\sim40'$, the correlation function is almost insensitive
  to $\Omega_0$. The dashed line shows the cluster normalisation from
  Eke et al. (1996).}
\label{om_sig_dependence}
\end{figure*}

Figure~\ref{om_sig_dependence} illustrates the dependence of
$w_{\kappa\delta}$ on $\Omega_0$ and $\sigma_8$, at three angular
scales, for fixed values of $h$ and $\bar b_\mathrm{R}$. For
measurements taken at a single angular scale only, the dependence on
$\Omega_0$ and $\sigma_8$ is degenerate. At small angular scales, for
instance, an increase in $\Omega_0$ can be compensated by a decrease
in $\sigma_8$ (cf.~the left panel of
Fig.~\ref{om_sig_dependence}). However, this degeneracy can be broken
if measurements at different angular scales are combined. At large
angular scales, for example, an increase in $\Omega_0$ would have to
be compensated by an increase in $\sigma_8$ (cf.~the right panel of
Fig.~\ref{om_sig_dependence}). Therefore, the shape of the correlation
function contains cosmological information, and it is thus important
to measure the correlation signal across a wide range of angular
scales. The dashed line in Fig.~\ref{om_sig_dependence} shows the
cluster normalisation constraint given by Eke et al.~(1996).

We note that there is an interesting angular scale around $40'$ where
the quasar-galaxy cross-correlation function becomes almost
independent of $\Omega_0$. The dependence of $w_{\kappa\delta}$ on
$\Omega_0$ and $\sigma_8$ can be understood as follows:

\begin{itemize}

\item From Eq.~(\ref{eq:7}), we see that the correlation function is
roughly proportional to $\Omega_0\times\int s\d
s\,P(s)\,\mathrm{J}_0(s\theta)$. At small angular scales, the Bessel
function is close to unity over a large range of wave numbers $s$, and
the result of the integration over $s$ is roughly proportional to
$\sigma_8^2$. Hence, we have $w\propto\Omega_0\,\sigma_8^2$, whatever
the value of $\Gamma$.

\item At large angular scales, the oscillations of the Bessel function
largely cancel the power at high wavenumbers, and thus the $s$
integration is only sensitive to the low-$s$ part of the power
spectrum. Since this rising part of the power spectrum decreases
rapidly with increasing $\Gamma$ (or equivalently increasing
$\Omega_0$) and keeping the normalisation constraint fixed, the
correlation amplitude decreases as $\Omega_0$ increases.

\item At intermediate angular scales, the various effects of changing
$\Omega_0$ cancel each other almost exactly, which leads to the
insensitivity of the correlation amplitude to $\Omega_0$ at angular
scales around $40'$.

\end{itemize}

The overall dependence of $w_{\kappa\delta}$ on $\sigma_8$ and
$\Omega_0$ reflects the physically motivated proportionality
$\Gamma\propto\Omega_0$, which is the simplest possible relation
between these parameters. More general cases will be taken into
account in Sect.~4.3.

Our main conclusions from this section are that the dependence of
$w_{\kappa\delta}$ on $\Omega_\Lambda$ is so weak that it can be
neglected, and the degeneracy between $\sigma_8$ and $\Omega_0$ can be
broken by measuring $w_{\kappa\delta}$ over a wide range of angular
scales between a few arc minutes and one degree. Assuming
$\Gamma=\Omega_0\,h$, $w_{\kappa\delta}$ becomes almost independent of
$\Omega_0$ for angular scales near $40'$. At that scale, the amplitude
of the correlation function is proportional to $\bar b\,\sigma_8^2$, and a
measurement of this product is in principle possible. We will address
this point in Sect.~4.2.

\section{Expected measurements}

We now evaluate how accurately the remaining cosmological parameters
can be constrained through measurements of quasar-galaxy
cross-correlation functions. For that purpose, we simulate
observations of quasar-galaxy cross-correlations expected to be
possible with the Sloan Digital Sky Survey data.

\subsection{Signal-to-noise estimate}

Since our goal is to constrain cosmological parameters from
quasar-galaxy cross-correlations, we now assess a strategy for
optimally using the data.  Within narrow, ring-shaped bins of radius
$\theta$ centred on
$N_\mathrm{Q}$ background QSOs, the total deviation of the galaxy
count from its mean is
\begin{equation}
  \sum_{i=1}^{N_\mathrm{Q}}
  \left(N_{\mathrm{G},i}-\bar{N}_\mathrm{G}\right)
  \approx N_\mathrm{Q}\,\bar{N}_\mathrm{G}\,
  w_{\mathrm QG}(\theta)\;,
\end{equation}
where $\bar{N}_\mathrm{G}$ is the mean galaxy count obtained from
randomly selected fields, and $N_{\mathrm{G},i}$ the galaxy count in
the ring centred on the $i$-th QSO. In a stack of such narrow,
ring-shaped bins, the signal-to-noise ratio of the detection will be
\begin{equation}
  \left(\frac{S}{N}\right) \approx 
  \sqrt{N_\mathrm{Q}\,\bar{N}_\mathrm{G}}\,
  w_\mathrm{QG}(\theta)\;.
\label{sn}
\end{equation}
As we saw before, the quasar-galaxy cross-correlation function
$w_\mathrm{QG}(\theta)$ is larger than the cross correlation
$w_{\kappa\delta}$ between convergence and projected density contrast
by a factor $2\,(\alpha-1)\,\bar b$. The value of the bias factor
$\bar b_\mathrm{R}$ is expected from simulations fall near
unity. However, the value of $(\alpha-1)$ depends on the quasar
sample.

The measured cumulative quasar number-count function strongly depends
on the selection criteria of the quasars. For our measurement, we
first need a cut at low redshift to exclude any overlap between the
QSOs and the galaxies. Then, for a given magnitude range, we need a
clean and complete sample. Since most of the SDSS quasars are
photometric, this implies cutting the sample in multicolour space for
optimising the selection criteria. We leave this level of detail aside
in this paper. We use the fitting function provided by Pei (1995) for
estimating the cumulative quasar number-counts as a function of
magnitude, i.e.~a broken power law with slopes of $\alpha\sim0.64$ for
QSOs fainter than $B\sim19$, and $\alpha\sim2.52$ for brighter QSOs.

Since faint QSOs have $\alpha<1$, they are negatively-biased by the
lensing magnification. Thus, taking all observed QSOs
straightforwardly into account would tend to reduce or even reverse
the correlation signal. However, we can separate the dependences of
$w_\mathrm{QG}(\theta)$ on angular scale (which contains the
cosmological information) and on quasar magnitude and thus keep all
information required to stack the correlation signal from all quasars
in a sample.

The Cauchy-Schwarz inequality provides optimal weights for quasars
with respect to magnitude. We find that the signal-to-noise ratio is
optimised by
\begin{eqnarray}
  \left(\frac{S}{N}\right)_{\rm opt} &\approx&
  \frac{\int\d m\,(\alpha(m)-1)^2\,N_Q(m)}
  {\int\d m\,|\alpha(m)-1|\,\sqrt{N_Q(m)}}\,
  \sqrt{\bar N_G}\;2\,\bar b_\mathrm{R}\,w_{\kappa\delta}(\theta)\;,
  \nonumber
\label{alpha_eff}
\end{eqnarray}
where $N_\mathrm{Q}(m)$ is the number of quasars in the magnitude bin
$[m,m+\d m]$, and $\alpha(m)$ is the corresponding slope of the number
counts. Therefore, we can consider the following effective correlation
function
\begin{equation}
  w_\mathrm{eff}(\theta)=
  2\,\bar b_\mathrm{R}\,(\alpha_\mathrm{eff}-1)\,
  w_{\kappa\delta}(\theta)\;,
\end{equation}
where $\alpha_\mathrm{eff}$ is an effective slope describing the
entire quasar sample,
\begin{equation}
  \alpha_\mathrm{eff}=1+\frac{1}{\sqrt{N_Q}}
  \frac{\int\d m\,(\alpha(m)-1)^2\,N_Q(m)}
  {\int\d m\,|\alpha(m)-1|\,\sqrt{N_Q(m)}}\;.
\end{equation}
The signal-to-noise ratio of the optimally weighted data can now
simply be calculated using $w_\mathrm{eff}$ instead of $w_\mathrm{QG}$
in Eq.~(\ref{sn}). The effective correlation function
$w_\mathrm{eff}(\theta)$ is related to an average {\em absolute\/}
deviation of the galaxy number counts in the vicinity of quasars from
their mean, weighted appropriately. In effect, each magnitude bin
gives a constructive signal to the cross-correlation, and the
weighting ensures a maximised signal-to-noise ratio. To get an idea of
a reasonable value for $\alpha_\mathrm{eff}$, we use Pei's (1995)
broken power-law fit for a sample of QSOs ranging in magnitude from
$m=16$ to $m=21$. Evaluating the integral in Eq.~(\ref{alpha_eff}), we
find that $\alpha_\mathrm{eff}=1.35$.

Some quasars have a magnitude $m$ for which $|\alpha(m)-1|\approx0$,
so they will only contribute noise to the final correlation
function. Those quasars are down-weighted by Eq.~(\ref{alpha_eff}),
but the selection can be further restricted by keeping only quasars
whose signal-to-noise ratio for the lensing effect exceeds a specified
threshold. The sample is then constrained by the condition
\begin{equation}
  |\alpha(m)-1|>\frac{1}{\sqrt{N_\mathrm{Q}\,\bar{N}_\mathrm{G}}\,
  2\,\bar b_\mathrm{R}\,w_{\kappa\delta}(\theta)}\;. 
\end{equation} 
The number of quasars satisfying this criterion is expected to be a
large fraction of the total available sample. We thus neglect the
sample restriction in the following, but it should be taken into
account in any measurement of $w_\mathrm{QG}$ from real data.

In the previous section, we emphasised the importance of measuring the
cross-correlation across a wide range of angular scales with a good
signal-to-noise ratio. Thus, the size of the ring-shaped bins around
quasars in which galaxies are counted must change with the ring radius
$\theta$. At large radii, wide bins are needed, and the earlier
narrow-bin calculation no longer holds. The general result for an
annular bin is
\begin{equation}
  \left(\frac{S}{N}\right)=\sqrt{\bar{n}_\mathrm{G}\,N_\mathrm{Q}}\,
  \frac{\int\d^2\phi\,w_\mathrm{eff}(\phi)\,
  [U_{\theta'}(\phi)-U_\theta(\phi)]}
  {\sqrt{\pi(\theta'^2-\theta^2)}}\;,
\label{sn2}
\end{equation}
where $U_\theta$ is a top-hat filter of radius $\theta$. This
integration over a disk replaces the zeroth-order Bessel function
$\mathrm{J}_0(s\theta)$ in Eq.~(\ref{eq:7}) by
$2\pi\theta\mathrm{J}_1(s\theta)/s$, and thus
\begin{eqnarray}
\label{J1eq}
  \int\d^2\phi\,w_{\kappa\delta}(\phi)\,U_\theta(\phi)&=&
  \int\d w\,\frac{p_\kappa(w)\,p_\delta(w)}{f_K^2(w)}\\
  &\times&
  \int \d s \,P_\delta\left(\frac{s}{f_K(w)},w\right)\,
  \theta\,\mathrm{J}_1(s\theta)\;.\nonumber
\end{eqnarray}

In order to adapt this model to a forthcoming application to the SDSS
data sample, we will use the following parameters:

\begin{itemize}

\item The galaxy density is
$\bar{n}_\mathrm{G}\simeq3\,\mbox{arcmin}^{-2}$ (Scranton et al.~2001)
for SDSS galaxies down to $r'=22$. It is shown that masking of stars
or bad seeing is important at that magnitude limit, and the usable
area of the survey is reduced by a factor 2/3.

\item The angular coverage is taken from $\theta=0.5'$ to $1^\circ$
and separated in 15 logarithmically spaced bins.  Below $0.5'$, the
shot noise becomes important due to the fairly low galaxy number
density. Above one degree, the amplitude of the correlation function
drops rapidly as shown in Fig.~\ref{shape}, and so does the signal of
the galaxy overdensity.

\item The effective logarithmic slope of the quasar number-count
function is set to $\alpha_\mathrm{eff}=1.35$.

\item The quasar number $N_\mathrm{Q}$ is essential for determining
the signal-to-noise ratio. Its value is continuously increasing with
the size of the survey and will reach its maximum in 2005. The SDSS
Early Data Release contains approximately 4000 spectroscopic quasars
and 9000 quasar candidates based on photometry alone. The expected
quasar number of the final survey is on the order of $10^5$ for
spectroscopically identified quasars, and can reach on the order of
$10^6$ for photometrically selected quasars (G.~Richards, private
communication). Furthermore, we need a quasar sample which is well
separated in redshift from the galaxy sample. For this purpose, we
will only use quasars with $z\ge1$. In order to account for these
numbers and for the reduction of the usable survey area due to
masking, we will consider an effective number of quasars
$N_\mathrm{Q}$ between $5\times10^3$ and $5\times10^5$.

Photometric redshifts will be available for all quasars. The amplitude
of $w_\mathrm{QG}$ is mainly sensitive to the mean redshift of the
considered quasar sample rather than its exact redshift
distribution. Therefore we can safely neglect any uncertainties in the
redshift estimates.

\end{itemize}

\subsection{Constraints on $\bar b(40')\,\sigma_8^2$}

\begin{figure}[ht]
  \includegraphics[width=\hsize]{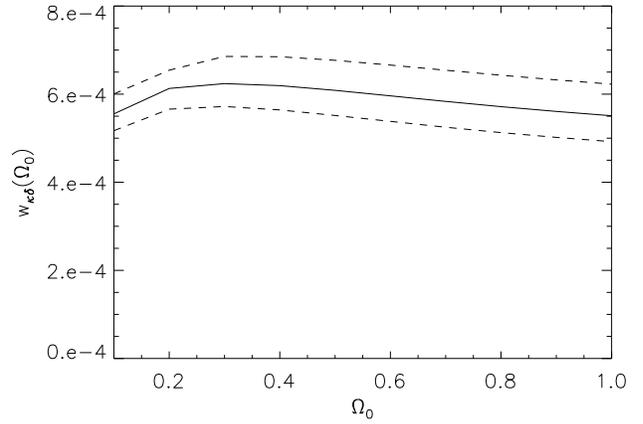}
\caption{$w_{\kappa\delta}(\Omega_0;\theta=40')$ is plotted for
  different values of $\Gamma$. The solid line is for
  $\Gamma=\Omega_0\,h$ with $h=0.72$, and the two dashed line are for
  $h=0.64$ and $h=0.80$, which are the 1-$\sigma$ confidence limits
  given by the Hubble Key Project. Here we have used $\sigma_8=1$.}
\label{sigma_dependence}
\end{figure}

We saw earlier that assuming $\Gamma=\Omega_0\,h$ (or similar
relations) causes the quasar-galaxy cross-correlation to be only
weakly sensitive to $\Omega_0$ near the angular scale of $\sim40'$
(corresponding to an effective scale of roughly $8\,{\rm
h}^{-1}_{50}\,{\rm Mpc}$), and this property can in principle be used
to further reduce the parameter space. At angular scales near $40'$,
the density perturbations probed by the correlation function are well
within the linear regime, where the amplitude of the power spectrum
simply scales $\propto\sigma_8^2$. At that scale, the correlation
amplitude scales $\propto b(40')\,\sigma_8^2$ to good accuracy, almost
independently of the density parameter $\Omega_0$ and the cosmological
constant $\Omega_\Lambda$. Given $\Omega_0$, the shape parameter of
the power spectrum is fixed by the Hubble constant. The product
$\bar b(40')\,\sigma_8^2$ can therefore be determined through the
correlation amplitude only up to the level of uncertainty reflecting
the uncertainty of the Hubble constant, and by the remaining weak
dependence on $\Omega_0$. Figure~\ref{sigma_dependence} shows how the
amplitude of the quasar-galaxy cross-correlation function near $40'$
changes with the Hubble constant and the density parameter
$\Omega_0$. The overall uncertainty is on the order of $20\%$.

The variation of $w_{\kappa\delta}$ with $\Omega_0$ can be reduced by
narrowing the angular range from which the data are taken, however at
the expense of the signal-to-noise ratio. Within an annulus with inner
radius $30'$ and outer radius $50'$, where the amplitude of
$w_\mathrm{QG}(\Omega_0)$ changes by less than about $20\%$, the
expected signal-to-noise ratio can be determined from Eq.~(\ref{sn2})
with the appropriate top-hat filters,
\begin{eqnarray}
  \left(\frac{S}{N}\right)&=&
  \sqrt{\bar{n}_\mathrm{G}\,N_\mathrm{Q}}\,
  \frac{\int\d^2\phi\,w_\mathrm{eff}(\phi)\,
  [U_{50'}(\phi)-U_{30'}(\phi)]}{\sqrt{\pi(50'^2-30'^2)}}
  \nonumber\\
  &\simeq& 9\,
  \left(\frac{N_Q}{10^5}\right)^{1/2}\,
  \left(\frac{\bar{n}_\mathrm{G}}{3}\right)^{1/2}\,
  \left(\frac{2(\alpha_\mathrm{eff}-1)\,b(40')}{0.7}\right)\;,
\end{eqnarray}
where we have used the number of spectroscopic quasars expected to be
provided by the SDSS. Using photometric quasars, this signal-to-noise
ratio might be improved. Therefore, the accuracy of a measurement of
$\bar b(40')\,\sigma_8^2$ through quasar-galaxy cross-correlations is
limited by the uncertainties in the Hubble constant and the density
parameter if $\Omega_0$ is treated as unknown. We emphasise the fact
that the only external information we use for this method is the
Hubble constant. Any additional constraint on $\Omega_0$ can
substantially increase the precision of the $\bar b(40')\,\sigma_8^2$
measurement.

\subsection{Parameter estimates: $\Omega_0$, $\sigma_8$ and $\bar
  b_\mathrm{R}$}

\begin{figure*}[ht]
\begin{center}
  \includegraphics[width=0.4\hsize]{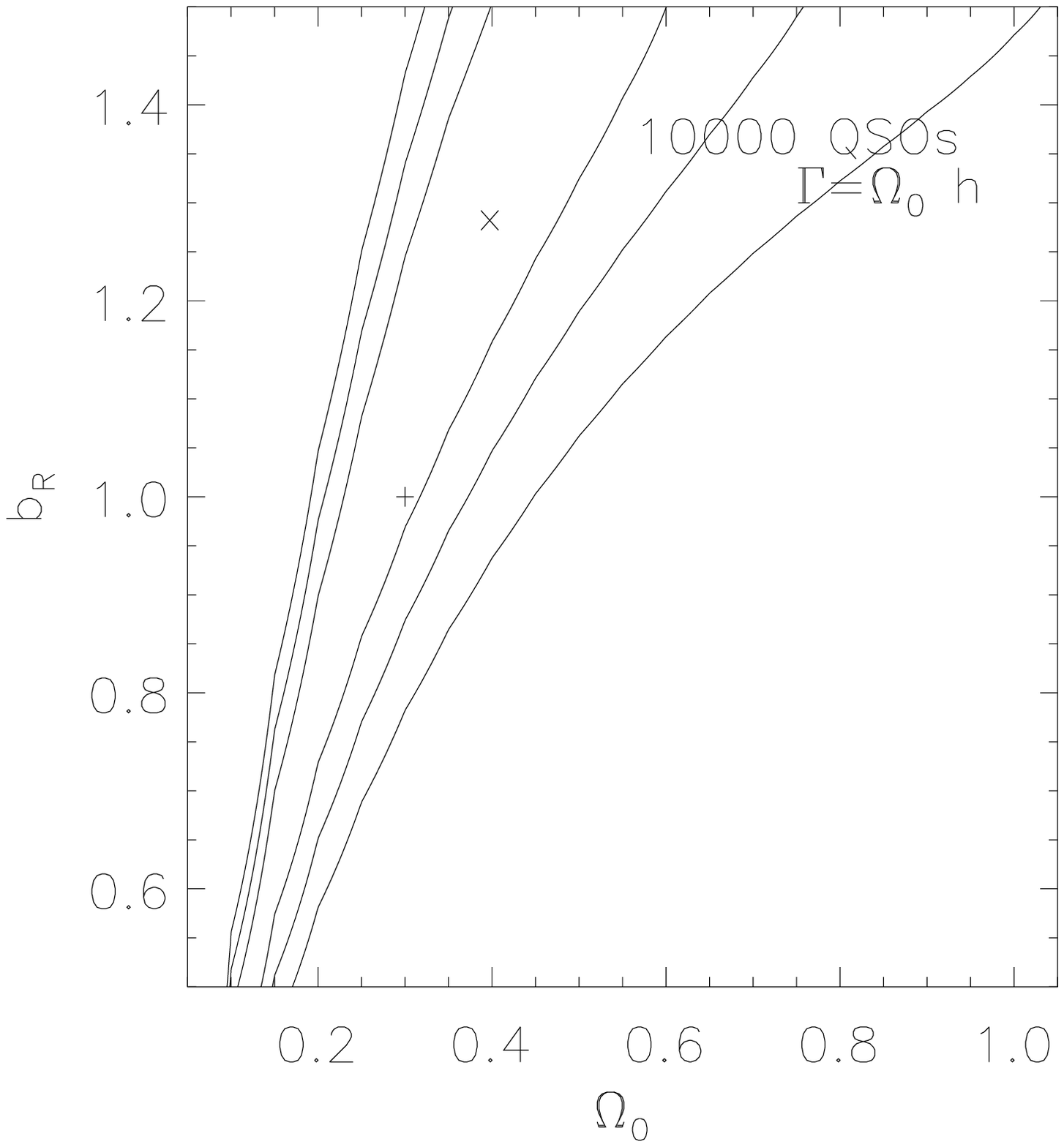}
\hspace{1cm}
  \includegraphics[width=0.4\hsize]{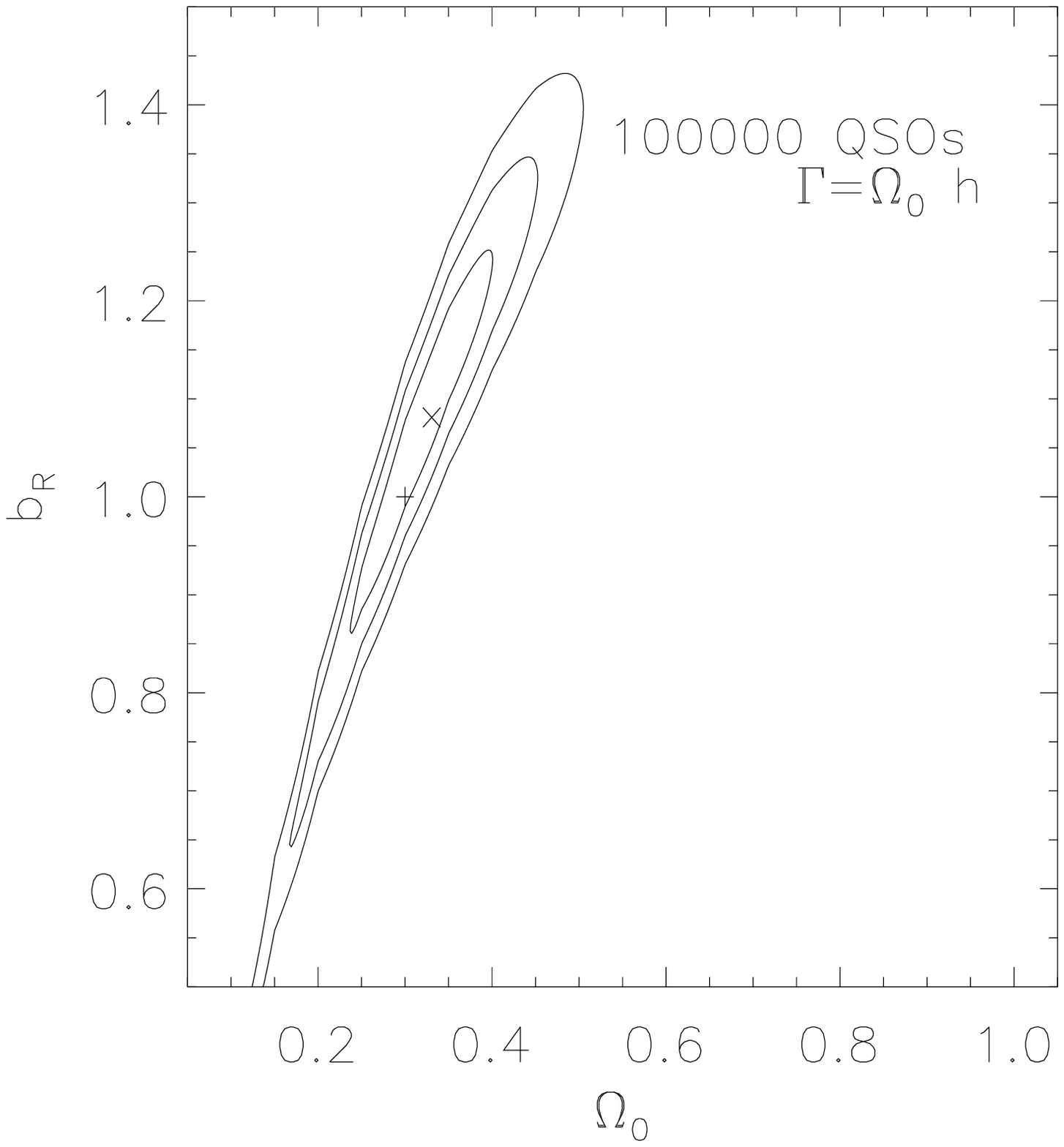}
\caption{Expected constraints on the parameters $\Omega_0$ and $\bar b$ for
  cluster normalised universes. The input parameters used to create
  the simulated observation are
  $(\Omega_0,\Omega_\Lambda,\sigma_8,\Gamma)=(0.3,0.7,1.0,0.21)$, and
  the bias was set to unity. The panels show the 1-, 2- and 3-$\sigma$
  confidence levels for quasar sample sizes of $10^4$, $10^5$, using
  the relation $\Gamma=\Omega_0\,h$. The symbol "$+$" shows the
  location of the input model, and the symbol "$\times$" shows the
  minimum of the $\chi^2$.}
\label{contours}
\end{center}
\end{figure*}

\begin{figure*}[!ht]
\begin{center}
  \includegraphics[width=0.4\hsize]{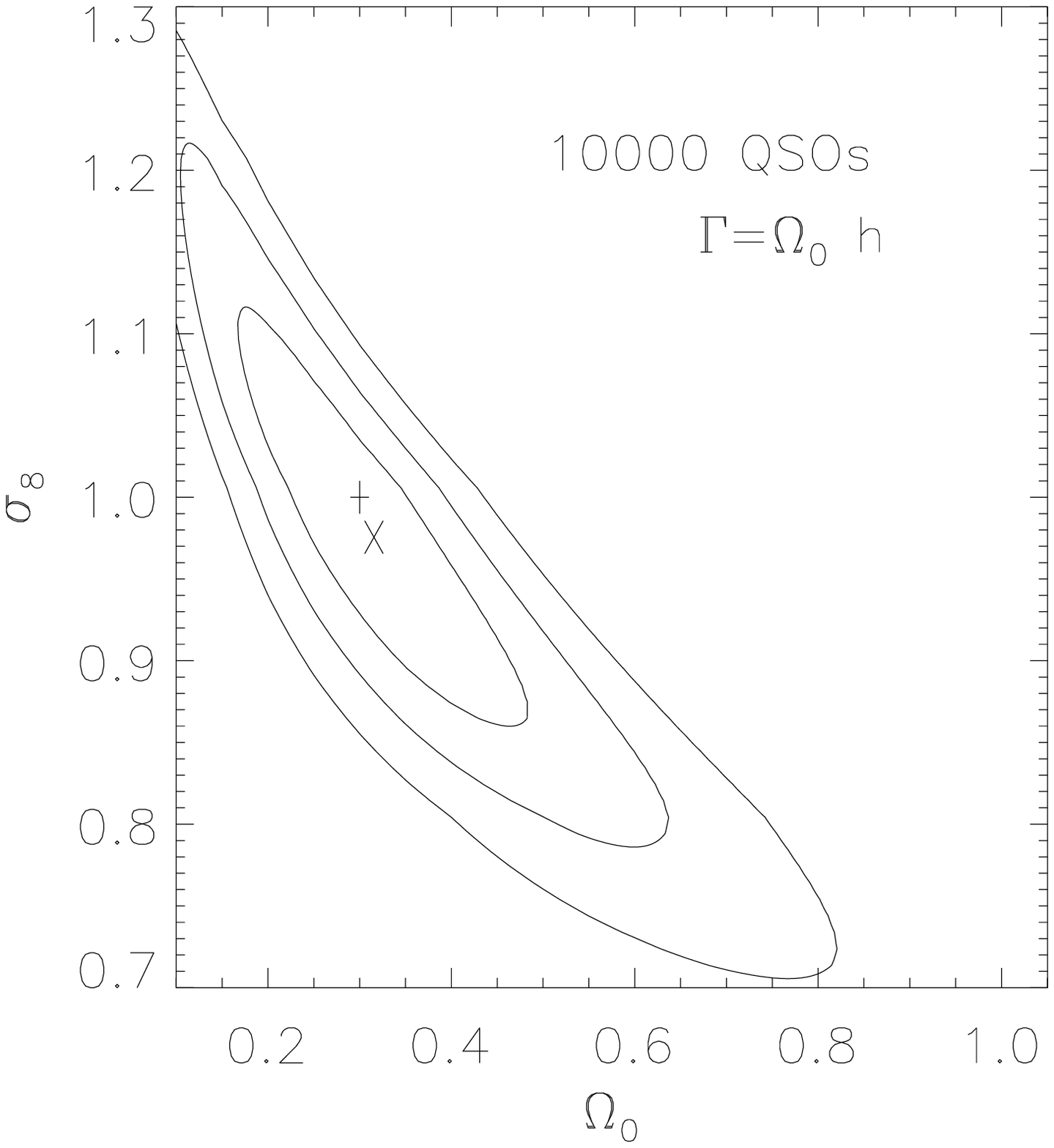}
\hspace{1cm}
  \includegraphics[width=0.4\hsize]{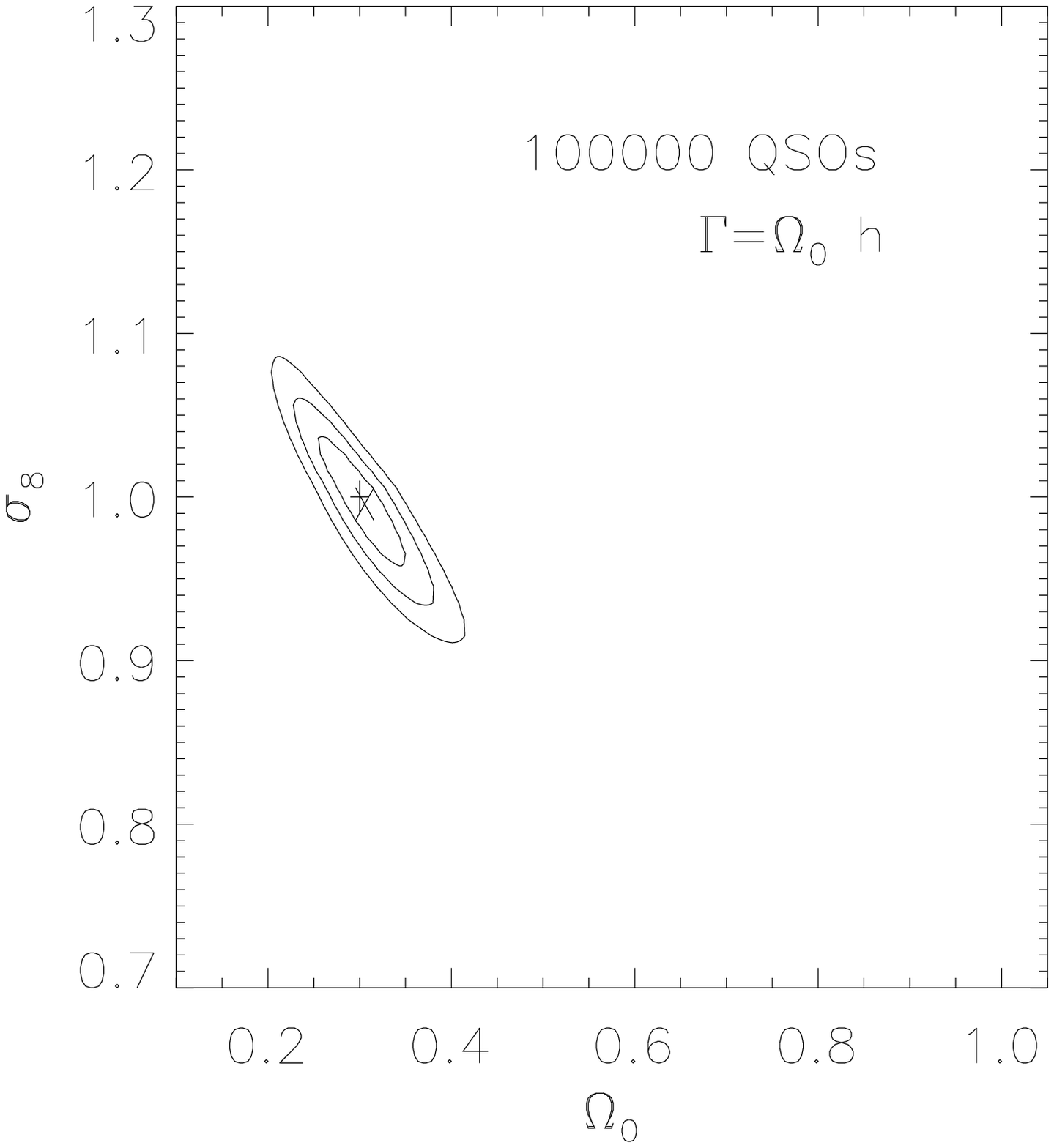}
\caption{Expected constraints on cosmological parameters. The contours
  show the 68\%, 90\%, and 95\% confidence levels. The two panels show
  results for quasar sample sizes of $10^4$, $10^5$. The parameters of
  the input model are
  $(\Omega_0,\Omega_\Lambda,\sigma_8,\Gamma)=(0.3,0.7,1.0,0.21)$, and
  $\bar b_\mathrm{R}=1$.}
\label{contours2}
\end{center}
\end{figure*}

We now investigate possible parameter estimates using the information
coming from all angular scales. Using Eqs.~(\ref{sn2}) and
(\ref{J1eq}), we can simulate measurements of $w_\mathrm{QG}(\theta)$
and then compute the uncertainty of the parameter constraints we wish
to extract from the data.

We saw above that the shape of the correlation function contains
important information on the cosmological parameters. For achieving
similar signal-to-noise ratios in all annular bins of the measurement,
we use logarithmically spaced annuli around the quasars. In order to
estimate the uncertainty of our parameters, it is worth noticing that
$10^5$ quasars spread evenly across the whole survey area are on
average separated by about $20'$. Given our 15 logarithmically spaced
bins covering the angular range introduced above, this means that the
errors in the three outer bins would be effectively correlated. The
others are only very weakly affected by possible correlations. Given
the low amplitude of the cross-correlation, the Poisson noise
dominates in our case. Hence, a simple $\chi^2$ of the difference
between the simulated data and the model enables us to reasonably
estimate the uncertainty of the parameters. However, it has to be kept
in mind that we are ignoring any angular variation in the bias, which
might contribute to the parameter uncertainties.

However, due to the large parameter space and some partial parameter
degeneracies, the $\chi^2$ minimisation yields allowed ranges for the
complete parameter set which are fairly wide compared to the existing
parameter constraints. An attempt to find constraints on the full
parameter space $(\Omega_0,\sigma_8,\Omega_\Lambda,h,\bar
b_\mathrm{R})$ is therefore not promising.  We have to include
additional information for better constraining our parameter set.

The prime advantage of the quasar-galaxy cross-correlations compared
to other methods for estimating the cosmological parameters is its
direct sensitivity to the galaxy bias factor. In addition, it is a
weak-lensing based method which can directly measure $\Omega_0$ and
provide an independent cross-check of results obtained from weak-shear
observations, as it does not rely on the assumption of intrinsically
uncorrelated galaxy ellipticities. The most accurate measurements of
$\Omega_0$ will be provided by CMB experiments. They rely on
information coming from photons emitted at the recombination epoch
($z\sim 1200$) and depend on the curvature of photon trajectories as
they propagate through the Universe. An independent measurement of
$\Omega_0$ using only low-redshift information probes the consistency
of the cosmological model.

For the following, we will use two external pieces of information. The
first is an estimate of the power spectrum normalisation
$\sigma_8$. For this, we use the cluster normalisation relation:
$\sigma_8=0.52\,\Omega_0^{-0.52+0.13\,\Omega_0}$ (Eke et al. 1996),
since there is no precise and direct measurement of $\sigma_8$. The
second is an estimate of the peak location of the power spectrum
$\Gamma$. We will use either a direct measurement, provided by galaxy
surveys, or a measurement of the Hubble constant $h$, if we adopt the
relation $\Gamma=\Omega_0\,h$. Finally, the remaining unknown
parameters of the problem are the density parameter $\Omega_0$, the
density parameter corresponding to the cosmological constant
$\Omega_\Lambda$, and the bias parameter $\bar b_\mathrm{R}$, of which
$\Omega_\Lambda$ can be ignored.

Other reductions of the parameter space are of course possible, but
this one is particularly attractive because it allows a measurement of
the galaxy bias and its dependences on scale, galaxy morphology,
magnitude etc. In what follows, our first fiducial model has
$\Gamma=\Omega_0\,h$, $h=0.72$, in agreement with recent constraints
from the Hubble Key Project (Freedman et al.~2001), and is cluster
normalised. We will later drop this relation between $\Gamma$ and
$\Omega_0$. Hence, we compute
\begin{equation}
  \chi^2(\Omega_0,b)=\sum_i\frac{\left[
    \delta N_\mathrm{M}(\theta_i;\Omega_0,\bar b_\mathrm{R})-
    \delta N_\mathrm{obs}(\theta_i;\Omega_0,\bar b_\mathrm{R,0})
  \right]^2}{\sigma_i^2}\;,
\end{equation}
where $\delta N_\mathrm{M}(\theta_i)$ and $\delta
N_\mathrm{obs}(\theta_i)$ are the overdensities of galaxies in bin
number $i$ expected in the model and detected in the simulated
observation, respectively, and $\sigma_i$ is the Poisson noise in that
bin.

The results are shown in Fig.~\ref{contours}. For this example, we
have assumed a cosmological model with $\Omega_0=0.3$,
$\Omega_\Lambda=0.7$ and a galaxy bias factor of unity. The contours
show the 68.3\%, 95.5\%, and 99.7\% confidence levels. Results for two
different quasar sample sizes are presented, namely $10^4$ and $10^5$
quasars, from left to right. We see that a large quasar sample is
required in order to at least confirm a low $\Omega_0$.

Previous quasar surveys listed only a few thousand objects, therefore
we expect the SDSS to provide us with the first quasar-galaxy
cross-correlation measurement with a reasonable signal-to-noise ratio
that allows the extraction of some cosmological information. The two
contour plots show a near-degeneracy between $\Omega_0$ and $\bar
b_\mathrm{R}$, for low $\Omega_0$. Indeed, by lowering $\Omega_0$ the
correlation function increases due to the cluster normalisation. For
low $\Omega_0$, mainly the amplitude of the correlation function
changes rather than its shape. Low values of the bias parameter can
compensate for this change and still allow a good fit. Thus, assuming
the relation $\Gamma=\Omega_0\,h$, quasar-galaxy correlations can
provide an estimate of $\Omega_0$, but only poorly constrain $\bar
b_\mathrm{R}$. Figure~\ref{contours} also shows that $\bar b$ can be
measured with good accuracy if the cosmological parameters are assumed
to be known; $\Delta\bar b_\mathrm{R}\approx15\%$ for $\Omega_0=0.3$.

As we discussed before, it is possible to probe other parameters with
the quasar-galaxy cross-correlation function. For example,
Fig.~\ref{contours2} shows what confidence intervals we can expect for
$\Omega_0$ and $\sigma_8$, assuming $h=0.72$ and $\bar
b_\mathrm{R}=1$. For instance, an independent measurement of the bias
can be obtained combining cosmic shear and galaxy counts provided
similar galaxy samples are considered (cf.~Schneider 1998 and van
Waerbeke 1998 for the theory, and Hoekstra et al.~2001b, in
preparation, for a measurement using a joint analyses of the Red
Cluster Survey and the VIRMOS-DESCART Survey). Using external
information on $\bar b_\mathrm{R}$, we see that interesting
constraints on $\Omega_0$ and $\sigma_8$ can be derived: we find
$\Delta\Omega_0\sim 10\%$ and $\Delta\sigma_8\sim5\%$.

\begin{figure*}[!ht]
\begin{center}
  \includegraphics[width=0.4\hsize]{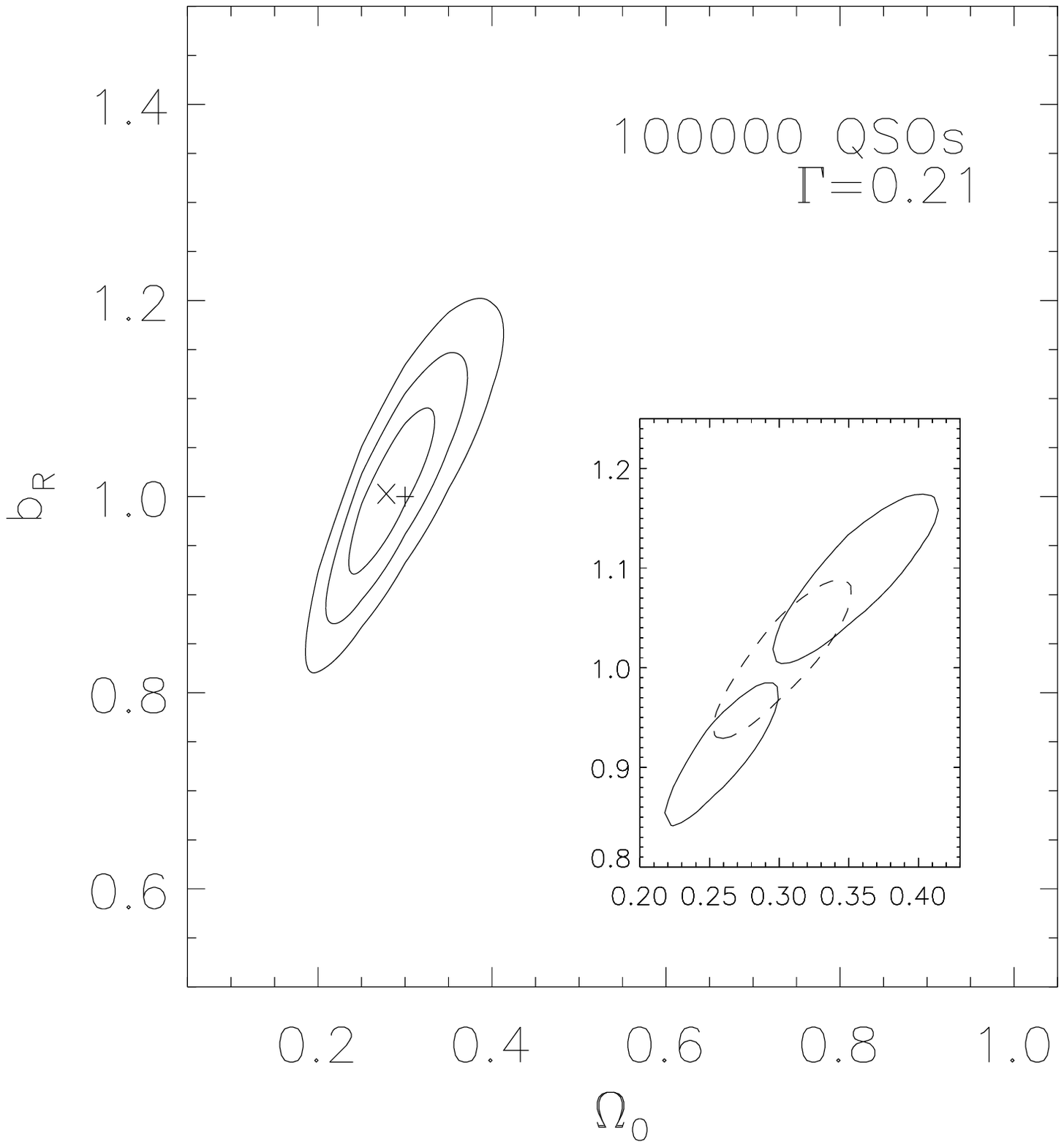}
\hspace{2cm}
  \includegraphics[width=0.4\hsize]{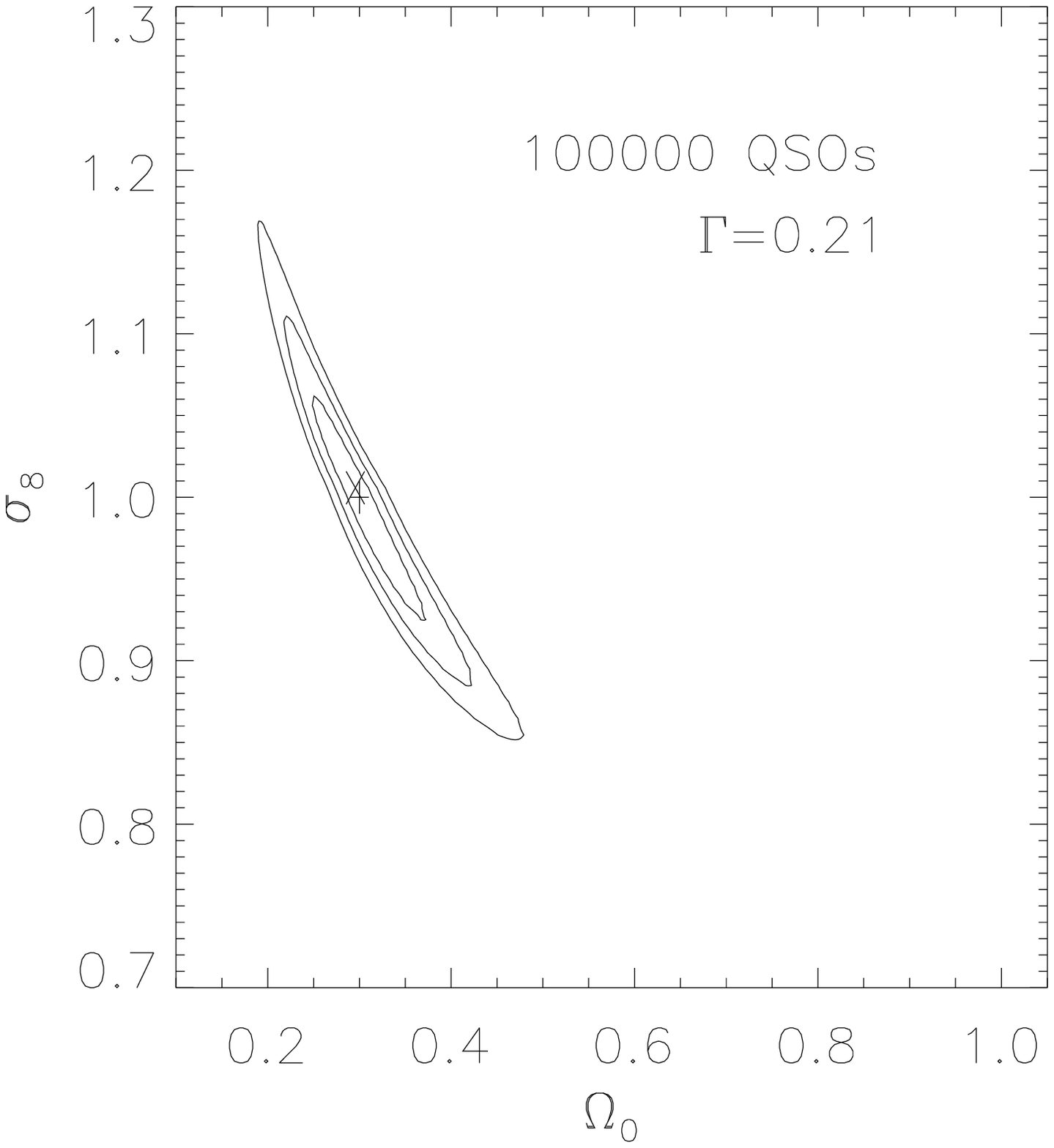}
\caption{Expected constraints on cosmological parameters. The contours
  show the 68\%, 90\%, and 95\% confidence levels, for a quasar sample
  size of $10^5$, considering the fixed value $\Gamma=0.21$. The
  parameters of the input model are
  $(\Omega_0,\Omega_\Lambda,\sigma_8,\Gamma)=(0.3,0.7,1.0,0.21)$, and
  $\bar b_\mathrm{R}=1$. The frame in the left panel shows how the
  1-$\sigma$ contour moves by considering $\Gamma=0.21$, but having
  $\Gamma_\mathrm{true}=0.19$ or $0.23$.\label{contours3}}
\end{center}
\end{figure*}

The validity of the assumed relation $\Gamma=\Omega_0\,h$ (or more
generalised relations taking baryons and neutrinos into account)
depends on the matter and radiation content of the Universe, which is
presently uncertain. For a more conservative approach, one can drop
this relation and use measurements for $\Gamma$. The two plots in
Figure~\ref{contours3} show the corresponding changes of the
confidence levels. In the $\Omega_0$-$\bar b_\mathrm{R}$ plane, we see
that we then get interesting constraints of the parameters. As we can
expect in the $\Omega_0$-$\sigma_8$ plane, the contours are not
significantly affected, and an interesting sensitivity on $\Omega_0$ and
$\sigma_8$ remains. Similar contour plots can be obtained from
cosmic-shear measurements (e.g.~Van Waerbeke et al.~2001) by assuming
a fixed value for $\Gamma$. As shown in Figure~\ref{contours3},
quasar-galaxy cross-correlations measured with the SDSS will soon
enable us to probe these parameters equally well as cosmic shear does
with current deep imaging surveys.

Regarding $\Omega_0$ and $\bar b_\mathrm{R}$, we saw from the
different contour plots that we achieve much higher sensitivity by
using external information on $\Gamma$.  Table~1 summarises the
uncertainties on $\Omega_0$ and $\bar b_\mathrm{R}$ we can achieve,
depending on the quasar sample size.  We can reasonably expect the
final usable quasar number to be at least of the order of $10^5$,
which means an independent measurement of $\Omega_0$ with an accuracy
of $\sim17\%$, and of $\bar b_\mathrm{R}$ with an accuracy of
$\sim10\%$, should be feasible.\\
Clearly, the final goal of using quasar-galaxy correlations is probing
the galaxy bias, but as these plots show, this method can also be used
for constraining cosmological parameters.

\begin{table}[ht]
\label{table_uncertainty}
\begin{center}
\begin{tabular}{|c|rr|}
\hline
  effective quasar number &
  $\Delta\Omega_0$ & $\Delta \bar b_\mathrm{R}$ \\
\hline
$1\times10^4$ & $0.14$ & $0.25$ \\
$5\times10^4$ & $0.07$ & $0.12$ \\
$1\times10^5$ & $0.05$ & $0.09$ \\
\hline
\end{tabular}
\end{center}
\caption{1-$\sigma$ uncertainties expected for $\Omega_0$ and $\bar
  b_\mathrm{R}$ as functions of the effective quasar number, using
  SDSS parameters and the input values $\Omega_0=0.3$ and $\bar
  b_\mathrm{R}=1$. Here, we consider a cluster normalised $\Lambda$CDM
  cosmology and a fixed value for $\Gamma$.}
\vspace{-.5cm}
\end{table}

The additional parameter $\Gamma$ is now known to approximately $20\%$
(Dodelson et al. 2001), and its accuracy is expected to greatly
improve with 
SDSS in the near future. However, this uncertainty adds to the
uncertainty of $\Omega_0$ and $\bar b_\mathrm{R}$. To see the
implications, the frame in the left panel of Fig.~\ref{contours3}
shows the effect on the 1-$\sigma$ confidence level of using different
values of $\Gamma$. The dashed line represents the 1-$\sigma$ contour
computed for the correct value of $\Gamma$ used for the simulations
($\Gamma=0.21$); the upper and lower contours are computed for
$\Gamma=0.23$ and $\Gamma=0.19$, respectively.  From this plot, we can
see that mistaking $\Gamma$ by $10\%$ will introduce a systematic of
about $16\%$ in $\bar b_\mathrm{R}$ and $10\%$ in $\Omega_0$.

\section{Conclusion and prospects}

Background quasars and foreground galaxies are cross-correlated on
angular scales between $\sim1'$ and $\sim1^\circ$ through the
magnification bias due to weak lensing by large-scale structures. The
effect probes the dark matter distribution in the Universe in much the
same way as measurements of cosmic shear do, but by counting objects
instead of measuring shapes, and thus without the crucial assumption
that the ellipticities of background galaxies are uncorrelated. In
addition, the lensing-induced QSO-galaxy correlations probe the
relation between the distributions of galaxies and dark matter.

The existence of statistically significant correlations between
foreground galaxies and background quasars has firmly been
established. Up to now, however, the quasar-galaxy correlation
function $w_\mathrm{QG}$ could not be reliably measured because
homogeneous, sufficiently deep, wide-field galaxy surveys were
missing. With the advent of such surveys, for which the Sloan Digital
Sky Survey (SDSS) is the prototypical example, a measurement of this
weak lensing effect with high signal-to-noise ratio is now 
within reach.

Using quasar-galaxy cross-correlations and assuming a linear and
deterministic bias, Ben\'\i tez \& Sanz (2000) showed how to measure
$\Omega_0/\bar b$ at a given angular scale from the ratio
$w_\mathrm{QG}(\theta)/w_\mathrm{GG}(\theta)$ between the
quasar-galaxy cross-correlation and the galaxy-galaxy autocorrelation
functions. However, their method discards the information contained in
the amplitude and angular shape of $w_\mathrm{QG}$. In this paper, we
investigated what kind of cosmological information can be expected
from an accurate measurement of $w_\mathrm{QG}$ using the SDSS
data. Our main assumptions were that

\begin{itemize}

\item galaxies are biased relative to the dark matter,
  possibly in a non-linear or stochastic way.
\item the shape parameter of the power spectrum either satisfies
  $\Gamma\propto\Omega_0\,h$, or can be fixed by the peak location in
  galaxy power spectra;
\item the power spectrum is normalised such that the local abundance
  of rich galaxy clusters is reproduced; and
\item the lensing magnification is weak. $|\mu-1|\ll1$.

\end{itemize}

Under these assumptions, $w_\mathrm{QG}$ depends in principle on the
density parameter $\Omega_0$, the cosmological constant
$\Omega_\Lambda$, the Hubble constant $h$ (or $\Gamma$), the
normalisation of the power spectrum $\sigma_8$, and the bias parameter
$\bar b$. This parameter range is too large to allow any accurate
cosmological constraints from the quasar-galaxy correlations
alone. However, the parameter range can safely be reduced. We showed
that:

\begin{enumerate}

\item the dependence of the quasar-galaxy cross-correlation function
  on the cosmological constant $\Omega_\Lambda$ is entirely negligible
  on scales above one arc minute;
\item assuming the relation $\Gamma=\Omega_0\,h$, we find an
  interesting angular range within which the amplitude of
  $w_\mathrm{QG}$ is insensitive to $\Omega_0$, and where its
  amplitude is simply proportional to $\bar b(40')\,\sigma_8^2$;
\item neglecting any relation between $\Gamma$ and the other
  parameters and using the cluster normalisation, the matter density
  and the bias parameter can be constrained with with $17\%$ and
  $10\%$ relative accuracy, respectively;
\item knowing the value of the bias parameter, it is possible to
  accurately measure of $\Omega_0$ and $\sigma_8$, namely with $17\%$
  and $7\%$ relative accuracy, respectively.

\end{enumerate}
We used a scale-independent bias, but models or measured
constraints on $\bar b(\theta)$ can be included into the calculation of
the cross-correlation function $w_\mathrm{QG}(\theta)$.

Our knowledge of the cosmological parameters will greatly improve over
the next few years, partly through the wide-field galaxy surveys
themselves, but most dramatically through the CMB
experiments. Therefore, the quasar-galaxy cross-correlations will be
less important for determinations of cosmological parameters, although
they will allow important consistency checks. However, the most
important contribution from a measurement of $w_\mathrm{QG}$ will be a
direct determination of the galaxy bias factor on a broad range of
scales, for different luminosities and morphological types of
galaxies.

Williams \& Irwin (1998) and Norman \& Williams (2000)
cross-correlated APM galaxies with LBQS and 1-Jansky quasars and
claimed significant galaxy overdensities around quasars on angular
scales of order one degree. As the amplitude of their correlations
cannot be explained in terms of weak lensing by large-scale structures
in a CDM universe, it is highly important to reexamine them.

Finally, the shape of $w_\mathrm{QG}$ depends on the shape of the
underlying dark-matter power spectrum, and a measurement of
$w_\mathrm{QG}$ will therefore test whether the CDM power spectrum
adequately describes the weak-lensing effects by large-scale
structure. This, of course, is also probed by cosmic shear, but only
under the assumption that intrinsic galaxy ellipticities are
uncorrelated. Measuring $w_\mathrm{QG}$ thus provides a useful,
independent, weak-lensing based method for directly constraining the
dark-matter power spectrum.

\begin{acknowledgements}

We would like to thank Yannick Mellier, Ludovic Van Waerbeke and Simon
White for useful discussions.

\end{acknowledgements}

\end{document}